%% file: main.tex
\documentclass[12pt]{article}
\usepackage{amsfonts}
\usepackage{amssymb}
\usepackage{fullpage}
\usepackage{amsmath}
\usepackage{rotating}
\usepackage{natbib}
\usepackage[doublespacing, nodisplayskipstretch]{setspace}
\usepackage{todonotes}
\usepackage{titlesec}
\usepackage{makecell}

\usepackage{tikz}
\usepackage{xcolor}
\usetikzlibrary{decorations.pathreplacing}
\usepackage[margin=1in, footskip=0.4in]{geometry}

\usepackage{graphicx}
\graphicspath{{../Code/outputs/}}

\usepackage[font=small,labelfont=bf]{caption}
\usepackage{subcaption}

\begin{document}
	

\title{Benchmarking Machine Learning Models to Predict Corporate Bankruptcy}
	
	
\author{ {\large Emmanuel Alanis\footnote{McCoy College of Business at Texas State University, \textit{Email:} \texttt{e.alanis@txstate.edu}}} 
		\and {\large Sudheer Chava\footnote{Scheller College of Business at Georgia Institute of Technology; \textit{Email:} \texttt{sudheer.chava@scheller.gatech.edu}}}  
		\and {\large Agam Shah\footnote{School of Computational Science \& Engineering, College of Computing at Georgia Institute of Technology; \textit{Email:} \texttt{ashah482@gatech.edu}}} }
\date{}

\maketitle
\thispagestyle{empty}

\doublespacing
\normalsize

\vspace{24pt}
	\begin{abstract}
		\noindent Using a comprehensive sample of $2,585$ bankruptcies from 1990 to 2019, we benchmark the performance of various machine learning models in predicting financial distress of publicly traded U.S. firms. We find that gradient boosted trees outperform other models in one-year-ahead forecasts. Variable permutation tests show that excess stock returns, idiosyncratic risk, and relative size are the more important variables for predictions. Textual features derived from corporate filings do not improve performance materially. In a credit competition model that accounts for the asymmetric cost of default misclassification, the survival random forest is able to capture large dollar profits.  
	\end{abstract}

\clearpage
\setcounter{page}{1}

\section{Introduction}
The risk of bankruptcy in a publicly traded firm is of major interest to shareholders, creditors, and employees. Prior literature has investigated the predictive performance of different forecasting models, mainly the discriminant analysis with accounting information \citep{altman1968financial}, the distance to default structural model \citep{bharath2008}, and the hazard model with accounting and market information \citep{shumway2001forecasting, chava2004bankruptcy}.

In this paper we investigate the benefits of applying high dimensional machine learning (ML) methods to bankruptcy prediction. We use a comprehensive sample of bankruptcies for U.S. publicly traded companies from 1969 to 2019 with financial, market, macro, and text based predictors. We study the performance of eight ML algorithms: the hazard model of \cite{shumway2001forecasting} and \cite{chava2004bankruptcy} enhanced with a penalty function (LASSO and Ridge), bagged trees (random forest and survival random forest), gradient boosted trees (XG Boost and LightGBM), and two specifications of neural networks (one shallower and one deeper).

ML algorithms seem to be well-suited to bankruptcy prediction since they are able to handle a large number of covariates and also excel at uncovering complicated nonlinear relationships, selecting important predictors in a data-driven approach without relying on researchers' prior beliefs \citep{hastie2009elements}.

We evaluate the out-of-sample performance of the algorithms with walk forward one-year-ahead forecasts of bankruptcy from 1990 to 2019. We assess performance by estimating the area under the curve (AUC) of the receiver operating characteristic (ROC) which measures the trade-off between type I and type II errors in the forecasts. 

Our main findings show that tree ensemble methods outperform others. In particular, the gradient boosted tree XG Boost achieves an AUC of 92\%. In addition, we find that once we include the distance to default measure, accounting ratios, and stock market information, predictors which capture industry and macroeconomic characteristics do not seem to improve the forecasting accuracy in a meaningful way. We interpret this result as consistent with the view that bankruptcy is an extreme company outcome and firm-level characteristics—in particular stock market information—may already incorporate the predictive impact of observable macroeconomic variables. \footnote{However, we note that \cite{duffie2009frailty} provide evidence that unobserved frailty has an impact on default intensities. It is possible that unobservable economy-wide covariates may matter for predicting clustering of defaults. We also stress that the loss given default may be significantly impacted by market wide factors, since recovery rates seems to vary with industry- and economy-wide cycles (Chava, Stefanescu and Turnbull, 2011). We do not analyze the clustering of defaults and variation in loss given default based on unobservable macroeconomic variables. Hence, we refrain from making definitive statements in this regard.}


We compare the performance of the algorithms during two pronounced crisis periods: the dot-com bubble of the turn of the century and the global financial crisis of 2007-2009. The performance of the models does not decrease significantly; however, the predictors powering model performance seem to differ between calm and turbulent periods in the economy. Stock market variables (the annual stock and idiosyncratic firm risk) are the most important predictors for most algorithms during non-crisis periods. During the dot-com bubble and the global financial crisis many more variables acquire relevance, in particular accounting ratios that measure the weight of liabilities in a company. 

In addition, we investigate if features derived from unstructured textual data increase forecasting performance. Recent literature has shown that adding text based measures can improve a variety of prediction tasks in the finance domain. \citet{nguyen2021multimodal} show that adding text-based features along with accounting variables improve predictions of a firm's credit rating. \citet{tetlock2008more} analyze how quantitative measures of text are useful in predicting individual firms' performance. In a hazard model of default, \citet{mayew2015md} find that linguistic information in corporate disclosure provides significant explanatory power in predicting whether a firm will go bankrupt three years prior to the event. 

We construct sentiment scores from annual company filings and analyze whether they increase the bankruptcy forecasting performance. We generate seven scores from corporate disclosure for positivity, negativity, litigiousness, uncertainty, readability, and sentiment. For sentiment measures, we use the lexicon based model VADER, and the state-of-the-art deep learning language model FinBERT \citep{araci2019finbert}.\footnote{See \cite{chava2020buzzwords}, \cite{chava2021managers}, and \cite{chava2022measuring} for a detailed discussion of the application of these models in finance.}

In the smaller sample with text measures---electronic filings are available starting in 1993---we find that the predictive power of our ML models does not increase in a significant way after including these text variables as predictors. Our findings suggest that these sentiment scores from prior literature may not be able to discern an event as rare and extreme as bankruptcy in ML algorithms.

While our main tests include a comprehensive set of predictors, in robustness analysis we show that a sparse ML model with only eight predictors (the market information model of \cite{shumway2001forecasting} and \cite{chava2004bankruptcy}, the distance to default measure of \cite{merton1974}, and FinBERT) can still provide very high performance. 

Finally, we use the credit competition model of \cite{agarwal2008comparing} to gauge the economic significance of our predictions. The consequences of default misclassification are probably not symmetric. The losses associated with incorrectly predicting a defaulting firm are much larger than the cost of an incorrect prediction for a firm that does not default. The stylized loan competition model makes explicit this asymmetric cost and allows us to assess which ML algorithm may have a better performance in a realistic setting. Our findings in this section suggest the survival random forest of \cite{ishwaran2008random} is a promising model for practice, as it is able to capture large dollar profits.

Our paper contributes to recent literature that studies the benefits of ML in bankruptcy prediction. \cite{tian2015variable} used the least absolute shrinkage and selector (LASSO) method to evaluate variable selection in the hazard model, while \cite{barboza2017machine} evaluate the forecasting accuracy of different ML algorithms using as predictors the accounting ratios of \cite{altman1968financial}.

\section{Data Description}

Our bankruptcy dataset extends the comprehensive sample of 
\cite{chava2004bankruptcy} and \cite{alanis2018shareholder} to include 
bankruptcies from 1969 to 2019 for U.S. publicly traded companies. We collect 
financial information for all U.S. publicly traded firms from Compustat with positive debt and stock market information from CRSP and we include in our sample every firm-year with available information. Our full sample comprises 131,261 firm-years and 2,585  instances of bankruptcy; our definition of bankruptcy includes both Chapter 11 reorganizations and Chapter 7 liquidations. 

Our dataset provides two advantages over the prior literature: it is a more 
extensive sample of bankruptcies (as a comparison, \cite{barboza2017machine} have a sample of 582 bankruptcies) and we allow our sample to be naturally skewed towards the non-event, this reduces the in-sample performance of our models but the out-of-sample performance is more indicative of their real life usefulness since publicly traded companies rarely default.

\begin{figure}[h!]
	\centering
	\caption{Number of Defaults by Year.}
	\includegraphics[scale=0.85]{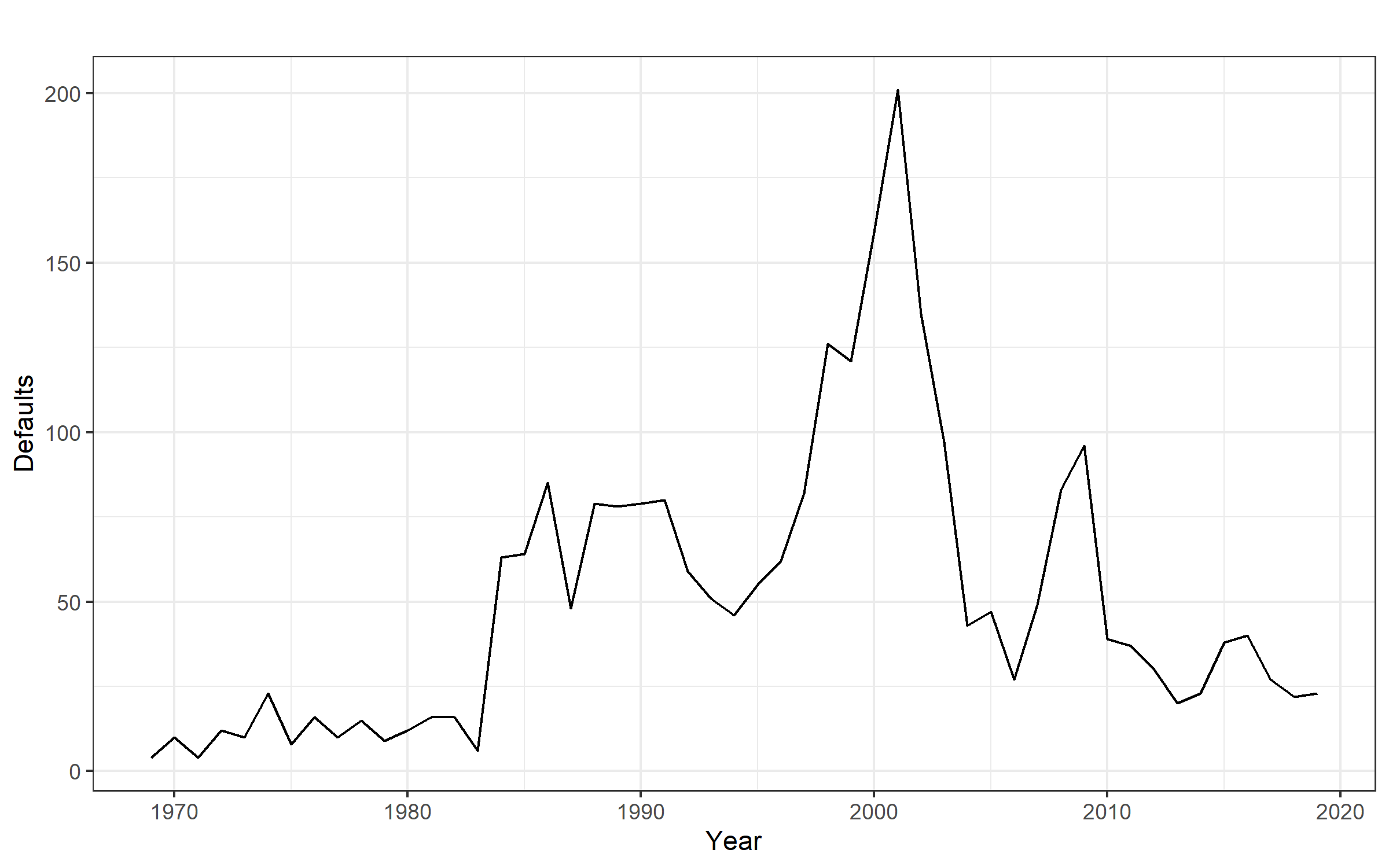}
	\label{fig:defyr}
\end{figure}

Figure \ref{fig:defyr} shows the time series of bankruptcies. The number of 
bankruptcies suffer a jump during the early 1980s, perhaps as result of the 
Bankruptcy Reform Act of 1978 which made it easier for companies to apply for Chapter 11 reorganization \citep{chavapurnanandam2010rfs}, and then is relatively stable with a large peak in 1999-2001 when the ``dot-com" bubble bursts and later another peak during the global financial crisis of 2007-2009. 

We can observe that default is not a common event for publicly traded firms, on average each year we have about 50 instances of bankruptcy and 2,520 firms that survive. This creates a challenge for any forecasting model of default since the target variable (default) is not balanced. 

\begin{figure}[h!]
	\centering
	\caption{Number of Defaults by Fama-French Industry.}
	\includegraphics[scale=0.85]{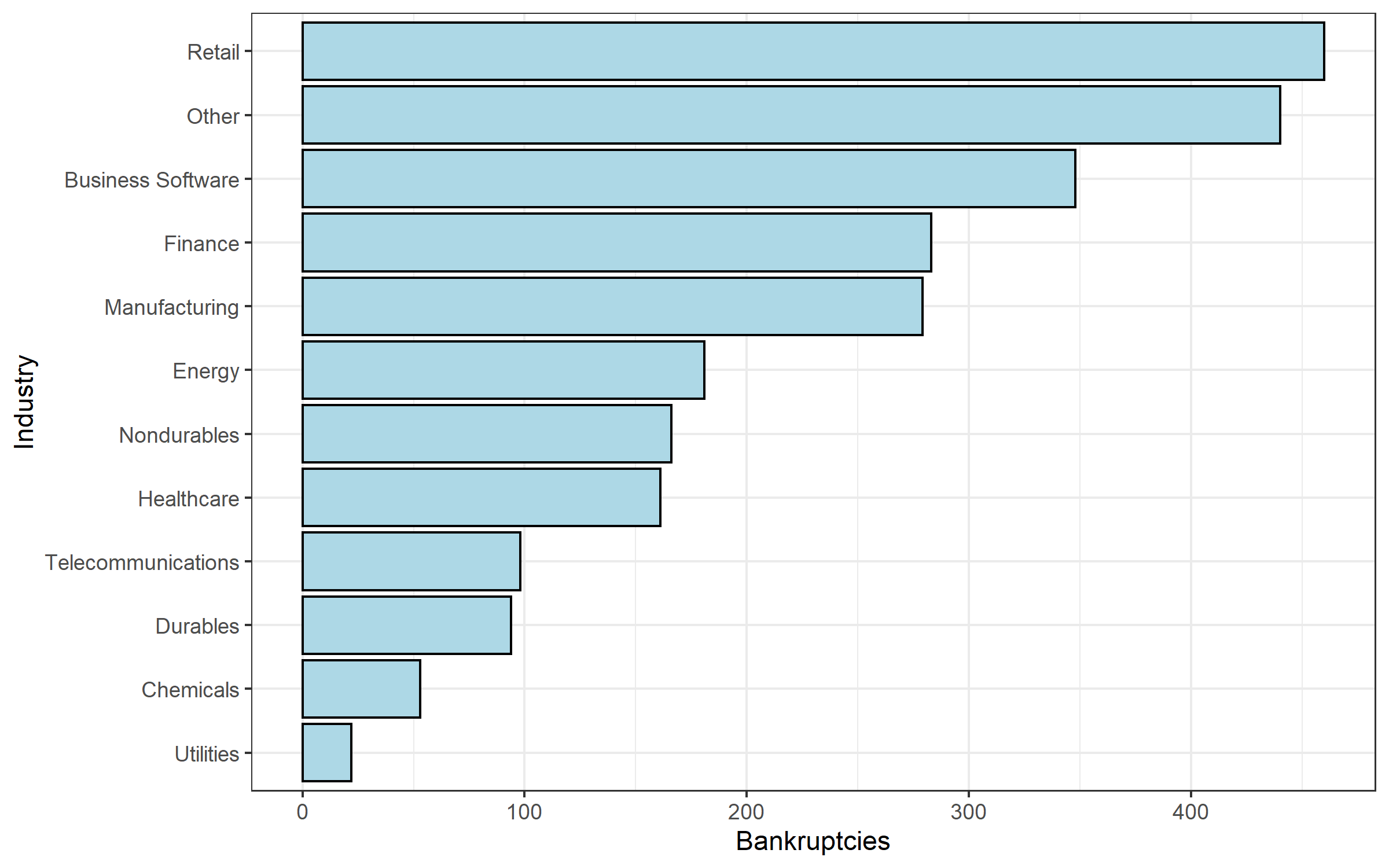}
	\label{fig:defind}
\end{figure}

Figure \ref{fig:defind} shows the distribution of bankruptcies across the 12 
Fama-French industry definitions.\footnote{We obtain the industry definitions from professor Ken French's website.} We can observe a strong effect of industry 
where some, such as Retail and Business Software, have many more bankruptcies 
than Utilities or Chemicals. The Other category, which has the second most 
bankruptcies, includes Transportation, Hospitality, and Entertainment, among 
others. The effect of industry on bankruptcy encompasses measures of regulation and competition; for instance, Utilities default rarely because they tend to be more tightly regulated and may not be free to choose their leverage level.

\subsection{Predictors}

Bankruptcy can be considered an idiosyncratic event for companies but, as 
Figures \ref{fig:defyr} and \ref{fig:defind} show, there is a nontrivial 
business cycle and industry component to it. We divide our bankruptcy 
predictors into three categories: company, industry, and macroeconomy. 

We follow \cite{shumway2001forecasting} in the way we construct our predictors. 
For every year $t$ we match financial information from the firm's annual 
statement that ends at least 6 months prior to the start of year $t$ and we use 
stock market information from year $t-1$. The lag in accounting information ensures this data is known to the market at the start of year $t$. We obtain our macroeconomic information for December of year $t-1$ from the FRED database at the Federal Reserve Bank of St. Louis.

We follow prior studies in the literature to construct company level predictors: 
\cite{altman1968financial}, \cite{shumway2001forecasting}, and 
\cite{chava2004bankruptcy}. We use the ratio of net income to assets ($ni/at$), 
total liabilities to assets ($tl/at$), net working capital to assets 
($nwc/at$), retained earnings to assets ($re/at$), the ratio of earnings before interest and taxes to assets ($ebit/at$), the ratio of market value of equity to total liabilities ($mkt/tl$), and the ratio of sales to assets ($sale/at$). We follow \cite{shumway2001forecasting} and impute missing values from financial 
statements with the last available observation. 

We also use the annual stock return in excess of the CRSP value-weighted index ($excess ~return$), the standard deviation of the residuals from the market model ($sigma$), and the log of the ratio of the market value of equity of the 
company to the total market value of the CRSP index ($relative~size$).

We also use the \emph{Distance to Default} measure from \cite{merton1974} which views a company's equity as a call option on its assets with the face value of debt as strike price. This measure shows how many standard deviations in its asset value a firm is away from the default boundary. \cite{bharath2008} show this measure has predictive ability for bankruptcy and we follow their methodology to construct this variable.

As industry variables, we use the Herfindahl-Hirschman index of sales 
concentration by the Fama-French 12 industry ($HH~Sales$), the median $sigma$ 
by industry-year ($sigma~industry$), and the industry-year median ratio of 
total liabilities to total assets ($tl/at~industry$). We choose these variables since $HH~Sales$ indicates the level of competition in the industry, while the other two provide information about the median level of risk in the industry.

Our macroeconomic predictors are variables that provide information about the current state of the economy and expectations about the future state of the economy. We use the difference in yield to maturity between the 10-year Treasury bond and the 1-year Treasury bill ($term~spread$), the difference in yield to maturity between Aaa and Baa rated bonds ($credit~spread$), and an indicator that equals 1 if the economy was in a recession in December of year $t-1$ according to the business cycle dates published by the National Bureau of Economics Research. We also include the annual inflation rate, the growth rate of GDP, the unemployment rate, and the annual growth in the industrial production index.

We augment firm level quantitative information with text-based features created from annual (10-K) Securities and Exchange Commission (SEC) company filings. We collect 10-K filings available on SEC-EDGAR\footnote{https://www.sec.gov/edgar/about} and extract clean text (by removing HTML and other non-text data) from the filings. We add four text measures by counting the number of positive, negative, litigious, and uncertainty words based on Loughran and McDonald dictionary \citep{loughran2011liability}. We normalize the word count by the total number of words in the filing and multiply them by 1,000, otherwise the quantities are extremely small. As readability measure, we calculate the Gunning-Fog index \citep{gunning1952technique} of the text.

In some cases, it is difficult to capture sentiment using the traditional bag-of-words approach. Hence, constructing a dictionary of words for sentiment measurement \citep[as used in][]{loughran2011liability} does not take advantage of word ordering and may give less accurate results. To overcome this issue, we calculate the sentiment of filings on sentence level using two approaches. First, we use the lexicon based VADER (Valence Aware Dictionary and sEntiment Reasoner) tool to calculate polarity (sentiment) of each sentence and then calculate a filing level measure by taking the mean of polarity of all sentences.

To take the advantage of recent advancements in natural language processing, we use pre-trained language model FinBERT \citep{araci2019finbert}, which is developed by further training a BERT-base \citep{devlin2018bert} model on financial data for sentiment analysis task. We use the fine-tuned model made available on Huggingface.\footnote{https://huggingface.co/ipuneetrathore/bert-base-cased-finetuned-finBERT} To calculate filing level sentiment measure based on FinBERT, we first calculate number of sentences with positive and negative sentiment and then use the following formula to get the sentiment measure.
\begin{equation}
    Sentiment_{FinBERT} = \frac{\# Positive Sentences - \# Negative Sentences}{\# Total Sentences}
\end{equation}

Table \ref{table:summary} contains summary statistics for our predictor variables. We separate our predictors into broad groups. We present first the \emph{Distance to Default} measure based on the structural model of \cite{merton1974}, then we present our group of accounting variables, then the set of variables that incorporate stock market information, then variables that capture industry variation, macroeconomic indicators, and finally text-based measures. This last group of variables are available only for the later part of our sample beginning in 1993, since prior there were no electronic filings in the SEC.

Some of our variables exhibit large variation, such as \emph{Net Income/Assets} and \emph{Mkt. Equity/Debt}, where standard deviations are orders of magnitude larger than the mean. The average firm-year in our sample has a negative \emph{Net Income/Assets} but this is due to some large negative outliers, and its median value is positive; the same is the case for the \emph{Net Income/Equity} ratio. In the text-based measures we observe that, on average, negative words roughly double positive ones; however, the FinBERT sentiment is slightly positive, which suggests positive sentences outweigh negative ones.

\begin{figure}[h!]
	\centering
	\caption{Correlation with Default.}
	\includegraphics[width=1.1\textwidth]{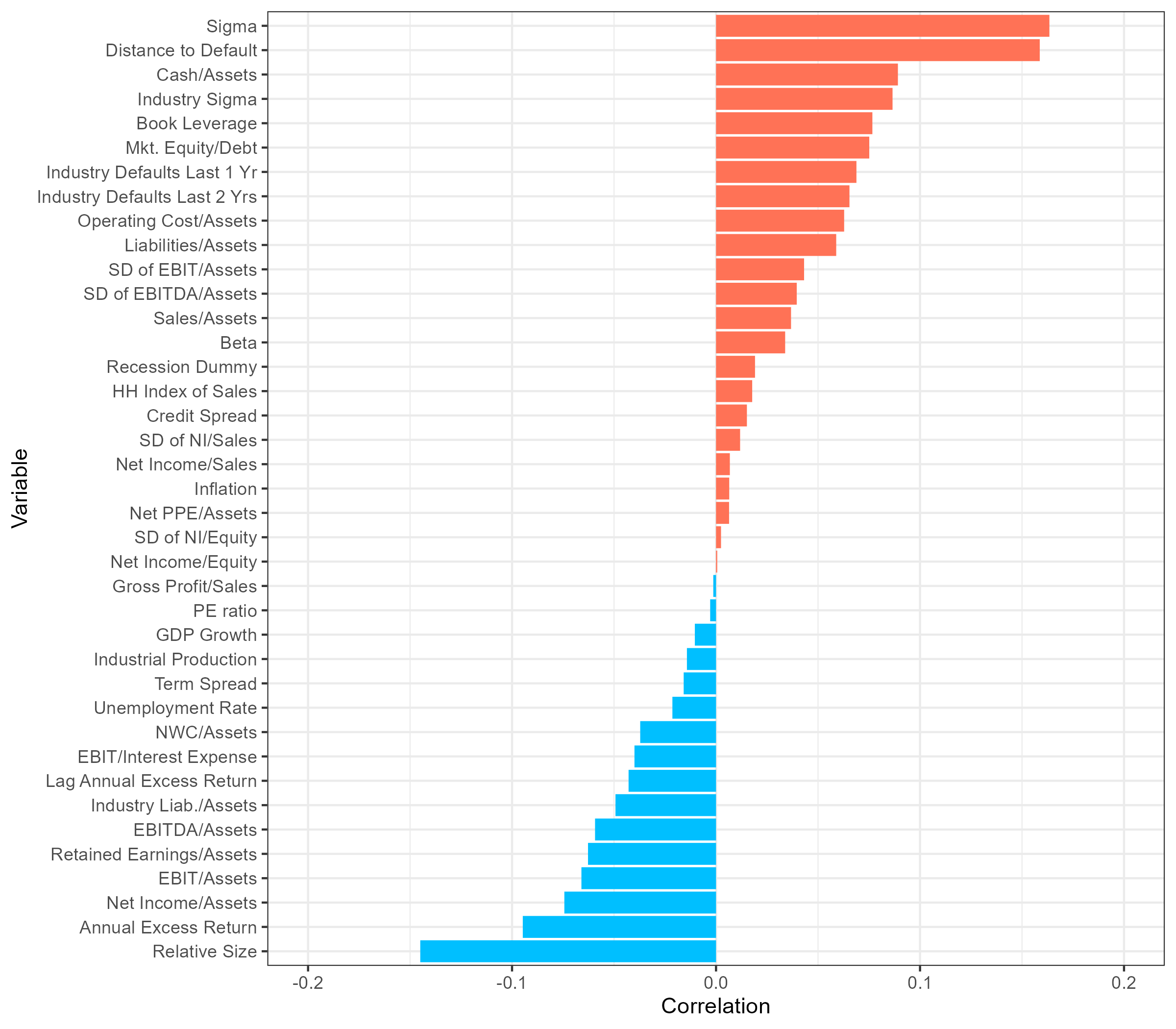}
	\label{fig:corr}
\end{figure}

Figure \ref{fig:corr} plots the Pearson correlation between all of our predictors and default. Machine learning algorithms are equipped to capture nonlinear interactions between the outcome variable and predictors, however, it is still instructive to analyze the linear correspondence that correlations indicate. 

Default is a complex event and no single variable has a strong correlation with it. The variables related to stock prices -- \emph{sigma}, \emph{annual excess return}, and \emph{relative size} -- are the ones that more tightly associated with default though correlations are less than 0.2 in magnitude.  This is similar to the finding in \cite{shumway2001forecasting} who shows that adding market information to a hazard model improves its performance over one with only accounting data. All of our macroeconomic variables -- \emph{credit spread}, \emph{term spread}, \emph{recession dummy}, \emph{inflation}, \emph{GDP growth}, \emph{unemployment rate}, and the \emph{industrial production} -- have small correlations, which suggests that on their own they may not be good predictors of default, which seems to be more associated with idiosyncratic (or firm level) variables.

The correlation between default and our text-based measures, not shown in Figure \ref{fig:corr} because of the different sample size, is also rather small. The FinBERT sentiment is the one with the biggest correlation with default at --0.06, followed by the negative word proportion from the \cite{loughran2011liability} dictionary at 0.05. 

\section{Methodology}

In this section we explain the different models we use to estimate bankruptcy predictions. We use eight machine learning algorithms which fall into four broad categories: penalized regression models, random forests, gradient boosted trees, and neural networks. We provide a high-level overview of these methods in the next subsections; for a textbook description of these algorithms see \cite{hastie2009elements} and 
\cite{aggarwal2018neural}.

\subsection{LASSO and Ridge}
In the context of default, the signal-to-noise ratio is low---bankruptcies 
remain a rare event among publicly traded firms---which increases the risk of 
fitting the noise in the estimation sample and rendering out-of-sample 
prediction useless.
 
\cite{shumway2001forecasting} proposed a logistic regression link between the 
bankruptcy indicator variable and time-varying covariates. The one-year-ahead 
default risk prediction is given by
\begin{equation}
	P(Y_{i,t+1} = 1 | Y_{i,t} = 0, X_{i,t}) = \frac{e^{\beta_0 + 
			\beta^{'}X_{i,t} } }{1 + e^{\beta_0 + \beta^{'}X_{i,t} }  }
	\label{eq:hazard}
\end{equation}
where $Y_{i,t}$ represents the bankruptcy indicator and $X_{i,t}$ are the 
predictors. \cite{shumway2001forecasting} and \cite{chava2004bankruptcy} estimated this regression model with a small set of accounting and market variables and showed it performs well relative to other regression models. We also estimate this model in order to benchmark our results against prior literature but given our large number of predictors we need a regularization tool.
 
The least absolute shrinkage and selection operator (LASSO) \citep{tibshirani1996regression} is a regularization 
tool that penalizes model complexity in an effort to avoid overfitting the 
sample data. We obtain the LASSO estimates for the hazard model of equation 
\eqref{eq:hazard} by minimizing the negative log likelihood function with a 
penalty weight $\lambda$ placed on the sum of the absolute value of covariate 
coefficient estimates (also called the $l_1$ penalty),
\begin{equation}
	\sum_i \left( -Y_{i,t+1} \left(\beta_0 + \beta^{'}X_{i,t}\right) + \log[1+ 
	\exp(\beta_0 + \beta^{'}X_{i,t}) ]\right) + \lambda \sum_{k=1}^{p}|\beta_k|
	\label{eq:lasso}
\end{equation}
where $p$ is the number of covariates. By adding the penalty term to the 
likelihood estimation the LASSO shrinks the coefficient estimates toward zero, 
while this worsens the in-sample fit of the model it can help improve the 
out-of-sample performance by avoiding overfitting.

The ridge regression \citep{hoerl1970ridge} is very similar to the LASSO except that it replaces the 
$l_1$ penalty with a $l_2$ cost. The penalty function for the ridge regression 
is $\lambda \sum^{p}_{k=1}(\beta_k^2)$ and also serves as a regularization 
tool. The main difference between the two estimates is that the ridge places 
little penalty on small values of $\beta$ but a rapidly increasing penalty on 
larger values, while the LASSO places a constant penalty on deviations from 
zero.

\subsection{Random Forest}
Random forest is a generalization of regression trees \citep{breiman1984classification}, which are fully 
nonparametric. A tree ``grows" in a sequence of steps. At each step, a 
new ``branch" sorts the data leftover from the preceding step into bins based 
on one of the predictor variables. The specific predictor variable upon which a 
branch is based and the value where it is split are chosen to 
minimize forecast error. 

A common problem of regression trees is their tendency to overfit the in-sample 
data giving poor out-of-sample predictions \citep{taddy2019business}. Random 
forests regularize decision trees through bagging, or model averaging. 
\cite{breiman2001random} set forth random forests by creating an ensemble, or 
forest, of trees where each one is fit to a random bootstrapped sample of the original 
data. In addition, only a random subsample of predictor variables is considered 
for each tree, which helps create more variation in each tree's forecast.

Bagging (short for bootstrap aggregation) independently fits decision trees 
onto random subsets of the training data. Each of the trees is simple and 
overfitted, but these errors from overfitting tend to be reduced when many 
trees are combined to create a forest.

The random forest method of \cite{breiman2001random} fits classification and 
regression decision trees (CART) on with-replacement samples of the data, and 
the resulting prediction is the average of the predictions from each tree in 
the bootstrap sample. While the prediction from each CART is likely fitting 
the noise in its estimation data, the overall average of bootstrapped 
samples will not be. On the contrary in order to take advantage of explainability, \cite{makwana4163283get} uses a single decision tree instead of random forest to derive rules which can help companies achieve investment grade rating. 

The random forest algorithm for a bootstrap size of $b=1 \dots B$ (this 
represents the number of trees in the forest) is to sample, with replacement, 
$n$ observations from the data. Fit a CART tree to 
each sample and get the prediction $\hat{y}_b$ from each tree. The prediction 
from the random forest is the average prediction of all the trees ($\hat{y}_F = \frac{1}{B} \sum_b \hat{y}_b$).\footnote{We use the classification random forest implementation of \cite{breiman2001random} where the Gini index is used as splitting rule.}

Since firm failure can be considered a censored event, we also estimate the 
survival random forest of \cite{ishwaran2008random} which extended the method 
of \cite{breiman2001random} to handle right-censored data. In this case, the 
response variable is a pair of values associated with each firm: the event time 
and the state of the firm (bankrupt or non-bankrupt).

\subsection{Gradient Boosting}
According to \cite{taddy2019business}, gradient boosting trees are the most 
important alternative class of tree-based estimators to random forests. 
Boosting is a strategy that sequentially fits an estimator to a training 
dataset and gives more weight to misclassified observations (or errors) in 
successive boosting rounds. Boosted trees iteratively estimate a sequence of 
shallow trees, each trained to predict the residuals of the previous tree. The 
final prediction is an accuracy-weighted average of the estimators. 

By giving more weight to successful estimators boosting can address bias, though it can increase the variance of predictions by over-weighting successful in-sample learners \citep{rasekhschaffe2019machine}. Most boosting methods include different options for regularization and early stopping in an attempt to stop the sequence of trees from learning the in-sample noise.

We use two gradient boosting tree algorithms, the XG Boost algorithm of 
\cite{chen2016xgboost} and the LightGBM method of \cite{ke2017lightgbm}. Both 
methods fit an ensemble of CART trees 
and then run successive rounds of boosting on them, they differ in the rules 
they follow to boost and split trees.

\subsection{Neural Network}
Neural networks (NNs) \citep{schmidhuber2015deep} represent a wide class of learning methods, arguably the most powerful---and least transparent---algorithm in ML 
\citep{gu2020empirical}. We focus on traditional “feed-forward” networks. These 
consist of an “input layer” of raw predictors, one or more “hidden layers” that 
interact and nonlinearly transform the predictors, and an “output layer” that 
aggregates hidden layers into an ultimate outcome prediction.

Figure \ref{fig:net} shows an illustrative example where the input layer 
consists of an intercept (also called ``bias") term $x_0$ and then other $D$ 
predictors. All of the predictors are passed (or feed-forward) to the first 
hidden layer, where each node of the layer applies---or activates---a 
potentially nonlinear transformation to the received inputs, weights them according to a parameter vector, and then feeds them forward to the next layer, and so 
on, until the output layer combines all signals into one prediction. 

\cite{gu2020empirical} argue that training a very deep (i.e. many hidden 
layers) NN is challenging due to the massive number of parameters and the 
potentially non-convex activation functions. We follow their strategy and 
instead of attempting to find the optimal network architecture we fix the 
network ex-ante. We consider two networks, one with three hidden layers (NN3) 
with 32, 16, and 8 nodes, respectively; our other choice is a network with five 
hidden layers (NN5) with 64, 32, 16, 8, and 4 nodes, respectively. For all nodes we 
use the rectified linear unit (ReLU) function which is defined as
\begin{equation}
	ReLU(x) = 
	\begin{cases}
		0 ,  & \text{if}~ x < 0\\
		x,              & \text{otherwise}
	\end{cases}
	\label{eq:relu}
\end{equation}

\subsection{Hyperparameter tuning}
All of the machine learning algorithms we consider need to be estimated and, in 
particular, have hyperparameters that should be ``tuned" so as to prevent an 
algorithm from overfitting the in-sample data. For example, two important 
hyperparameters that need to be tuned in a random forest are the number of 
parallel trees and the maximum depth allowed for each tree.

The common approach in the literature is to select tuning
parameters adaptively from the data in a validation sample. In particular, we
divide our sample into three disjoint time periods that maintain the temporal
ordering of the data: training, validation, and out-of-sample testing. 

\begin{figure}[h!]
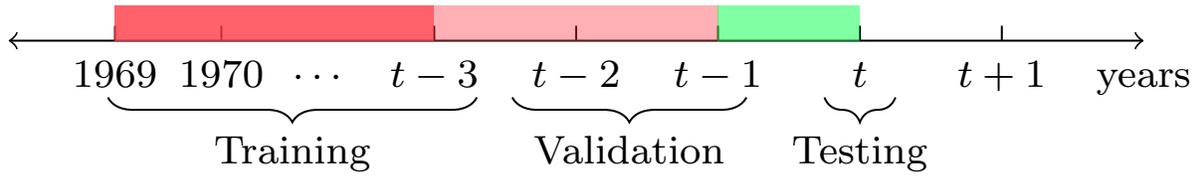

	\centering
	\caption{Expanding Window for Training/Validation/Testing.}
	\include{timeline}
	\label{fig:timeline}
\end{figure}

Figure \ref{fig:timeline} presents the way we split our observations for every 
year in our sample for each ML method. In order to forecast default for year \textit{t} in our sample, 
we first estimate a model subject to a specific set of tuning 
parameter values on a training sample which goes from 1969 (first year of our sample) to year $t-3$. In 
a second step we evaluate the set 
of hyperparameters by creating forecasts and computing 
the receiver operating characteristic area under the curve (AUC) in a 2 year validation sample of years $t-2$ and $t-1$. 
We iterate over a grid of hyperparameters in the first and second step and then select the set that minimizes the AUC in the validation sample. In a third 
step, we get the optimized model's prediction for year $t$, which was 
neither used for estimation nor tuning, so this is truly out-of-sample and 
provides an honest evaluation of a method's predictive performance.

As shown in Panel B of Figure \ref{fig:timeline}, to forecast year $t+1$ we expand the training window by one year to now also include year $t-2$, we still use two years for validation (years $t-1$ and $t$), and we forecast year $t+1$. Thus, we follow an expanding window on our training set and always use the two prior years for validation.

\subsection{Model evaluation}
For each model we predict the probability of bankruptcy in one year and we 
evaluate their performance by computing their sensitivity and specificity. The 
former is the probability of predicting default given that the true outcome 
is default; the latter is the probability of predicting non-bankruptcy when the 
true outcome is non-bankruptcy. Equivalently, sensitivity is type-I error and 
specificity is type-II error.

The receiver operating characteristic is a commonly used tool to summarize the tradeoff between sensitivity and specificity 
\citep{hastie2009elements}. It is a plot of the sensitivity versus 
1-specificity as we vary the probability cutoff to classify a prediction as 
bankruptcy or non-bankruptcy. We also evaluate model performance by computing the area under the curve (AUC) of the ROC, which measures performance across all possible classification thresholds. A model that is always correct has an AUC of 1.00 and a model which is just a random prediction has AUC of 0.5.


\section{Results}

We use the years from 1969 to 1989 as our initial training and validation 
period, then we forecast out-of-sample for 1990. We always conduct 
one-year-ahead forecasts by increasing our estimation sample by one year, as 
shown in Figure \ref{fig:timeline}. Our final out-of-sample forecast comprises 
the 30 years from 1990 until 2019.

Our baseline estimation only includes the \emph{Distance to Default} and the 12 
Fama-French industry indicators as predictors. The \emph{Distance to Default} 
has been examined in prior literature and under the structural model of 
\cite{merton1974} it captures all relevant information to predict bankruptcy. 
We add groups of predictors to this baseline specification one at a time to 
evaluate the impact of additional information. 

\begin{figure}[h!]
	\centering
	\caption{Receiver Operating Characteristic (ROC) plot.}
	\includegraphics[width=1.1\textwidth]{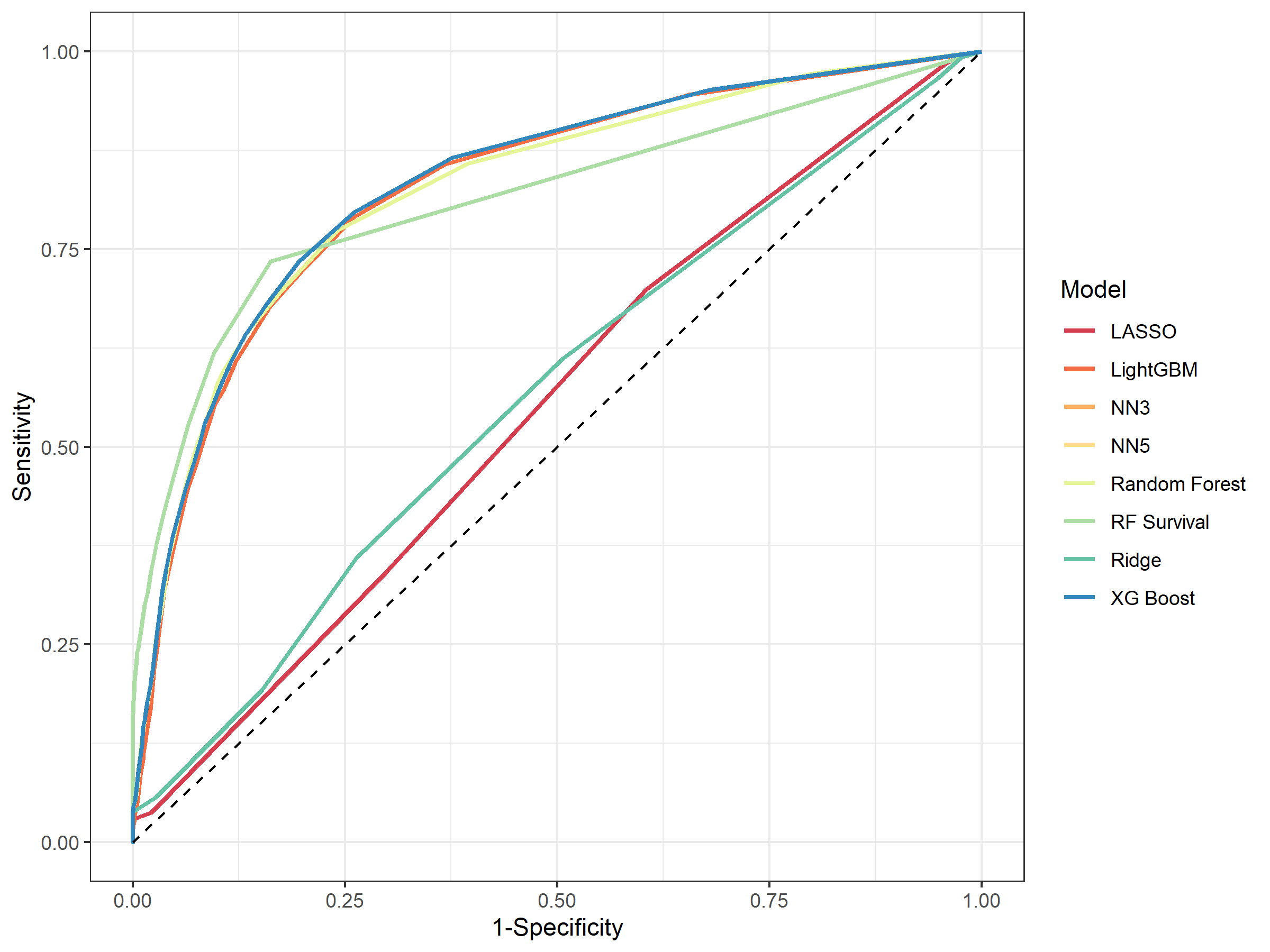}
	\label{fig:roc}
\end{figure}

Figure \ref{fig:roc} plots the ROC curves for our different models when the 
\emph{Distance to Default} and the industry indicators are the only predictors. 
The 45 degree line is the naive benchmark of a pure random forecast. A model 
with perfect 
forecast would have a 90 degree angle where it climbs the y-axis and then goes 
horizontal at 1. A model with a good fit will force the curve to the top-left 
corner, thus in a ROC plot the best models are closer to the top-left corner. 

We can see the LASSO and Ridge regressions are not very good in this 
instance. This is somewhat expected since these algorithms are most useful in 
high dimensional settings where they force sparsity in the specification, but 
when we have only a few predictors their usefulness is not maximized. The rest 
of the algorithms are all closely bunched and it becomes difficult to 
differentiate models, for this reason we compute the AUC, 
which is the integral under each ROC curve, this way we can compare models 
using a numerical measure.

Panel A of Table \ref{table:auc} shows the AUC computed for all models. The 
first column only includes the industry indicators and the \emph{Distance to 
Default} as predictors, the tree ensemble algorithms produce the best forecasts 
with an AUC of 0.83. Even though 
the number of predictors is very limited in this case, decision trees have the 
ability to uncover nonlinear relationships and it seems very valuable in this 
case.

The next column in Panel A adds the group of accounting variables to the set of 
predictors. We see the AUC increases for all models and the best performing one 
is the gradient boosted trees XG Boost which achieves an AUC of 0.92. The rest 
of the columns add more 
predictors, to the point where in the last column we use all of the predictors. 
As we compare across columns, it seems the industry and macroeconomic variables 
do not add much to the forecasting performance of the algorithms. The AUCs for 
most models plateaus after the \emph{Distance to Default}, accounting, and 
market variables are included. Firm-level characteristics, rather than market 
wide variables, seem more important for default prediction. 

Previous research has shown that defaults can cluster due to market wide variables \citep{duffie2009frailty, chava2011modeling} and our Figure \ref{fig:defyr} shows a business cycle component to the number of aggregate bankruptcies. However, our results suggest that after accounting for firm-level characteristics, market wide factors do not exhibit much predictive power with ML algorithms. It seems plausible that macroeconomic variables are important for aggregate bankruptcies, but firm-level characteristics already reflect their impact at the company level, and thus the specific default of a company depends on its own conditions.

Machine learning models rarely lend themselves to causal inference; 
nevertheless, it is possible 
to examine which predictors are powering the performance of the algorithms. We 
compute a permutation importance score for each predictor in our models. The 
permutation importance is defined as the decrease in the performance 
measure---the AUC in our case---when a single predictor is randomly shuffled in 
its value. This 
procedure breaks the relationship between the predictor variables and the 
forecasted variable, thus the change in AUC is indicative of how much the model 
depends on the predictor variable.

\begin{figure}[h!]
	\centering
	
	\caption{Variable Importance: Full Sample.}
	\includegraphics[width=1.1\textwidth]{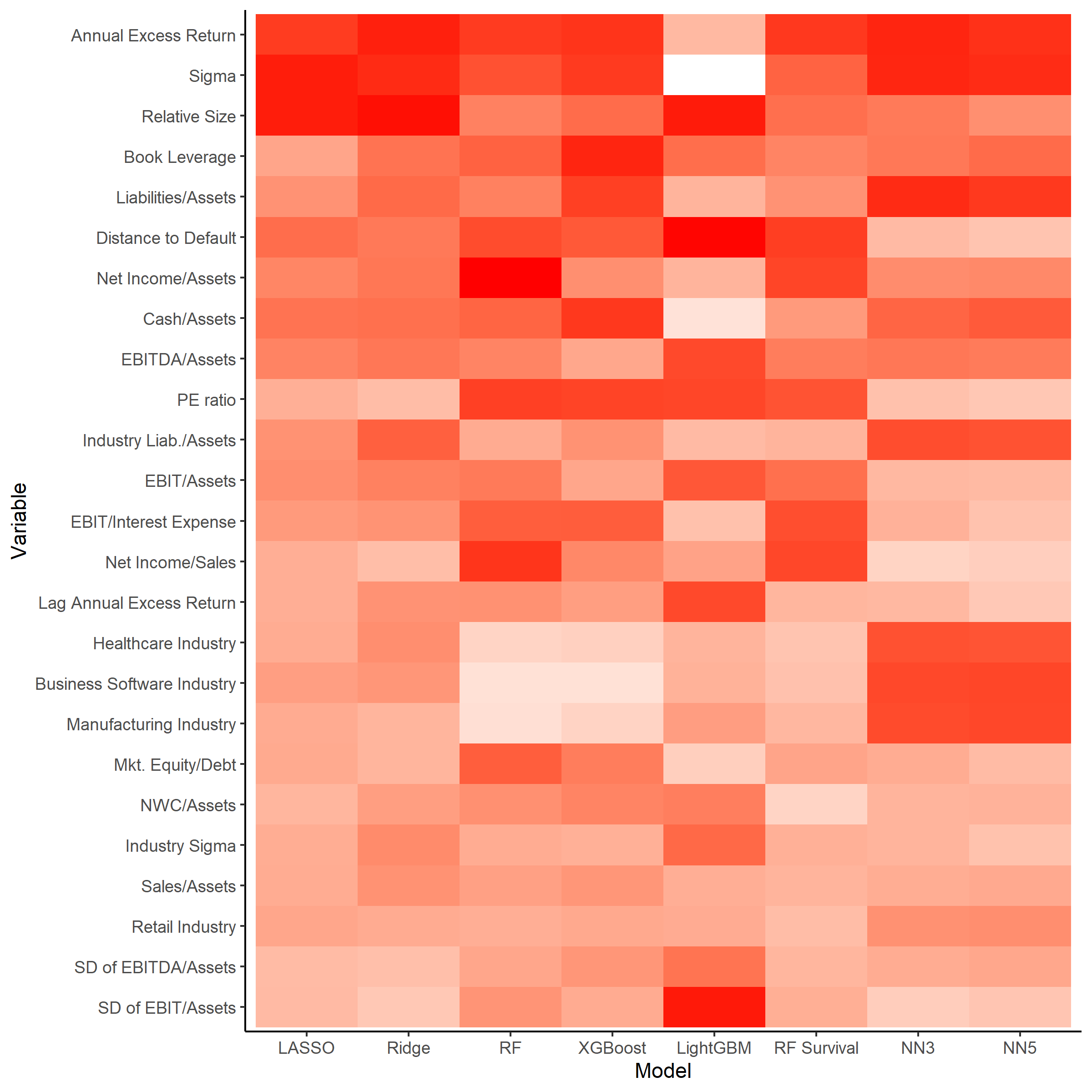}
	\label{fig:vimp}
\end{figure}

For each of our algorithms we rank the permutation score for all of our 
predictors and present the top 25 in Figure \ref{fig:vimp}. Predictor 
variables are ordered so that the ones with the best overall sum of rank are at 
the top and those with the lowest overall rank are at the bottom. The color 
gradient within each column shows the model-specific ranking of characteristics 
from least to most important (lightest to darkest red).

Most models agree the most relevant predictors are market driven variables: the 
\emph{Annual Excess Return}, \emph{Sigma}, and the \emph{Relative Size}. The LASSO and Ridge algorithms---which favor sparsity---do not find much impact from other variables. The tree ensemble methods tend to agree between 
themselves on the ranking, with the notable exception of LightGBM which does 
not seem to find the \emph{Annual Excess Return} and \emph{Sigma} relevant and 
instead the \emph{Distance to Default} more important. It is noteworthy that 
only the neural networks (NN3 and NN5) find value in the industry indicators, 
all other models do not find them all that relevant. Some of the accounting 
variables are relevant in particular models---such as \emph{Net Income/Assets} 
and \emph{Net Income/Sales} for the random forest, \emph{Book Leverage} for XG 
Boost, and the standard deviation of \emph{EBIT/Assets} for LightGBM---but by 
and large there is no agreement between the models that accounting variables 
are important. This is consistent with \cite{shumway2001forecasting} who found 
that market variables tend to dominate accounting ratios in his hazard model of 
default, which the author attributes to market variables being forward looking while accounting variables reflect the recent past.

\subsection{Crisis periods}

Figure \ref{fig:defyr} shows that the number of defaults spike during and 
shortly after crisis periods, such as the burst of the dot-com bubble and the 
global financial crisis. We evaluate the performance of the machine learning 
models for predicting default during these two crisis periods. 

Panel B of Table \ref{table:auc} shows the AUC for all models when forecasting 
bankruptcy for the years 1999 to 2001, when the dot-com bubble burst. We 
observe a similar pattern to the full sample on Panel A, where tree ensembles 
have moderate performance even when only including the industry indicators and 
\emph{Distance to Default} as predictors. The model performance improves as 
more predictors are added, but, again, the industry and macro indicators do not 
seem to add much value. The tree ensemble methods, as in the full sample, have 
the best performance and again the boosted tree algorithm of XG Boost has the 
best AUC.

Panel C shows the forecasting performance for years 2007 to 2009, around the 
global financial crisis. We observe a similar pattern to Panels A and B, where 
the gains of including industry and macro variables are marginal at best and 
the tree models perform well, though in this case the 5 layer neural network 
has an even better performance.

\begin{figure}[h!]
	\centering
	
	\caption{Variable Importance: Dot-com Years.}
	\includegraphics[width=1.1\textwidth]{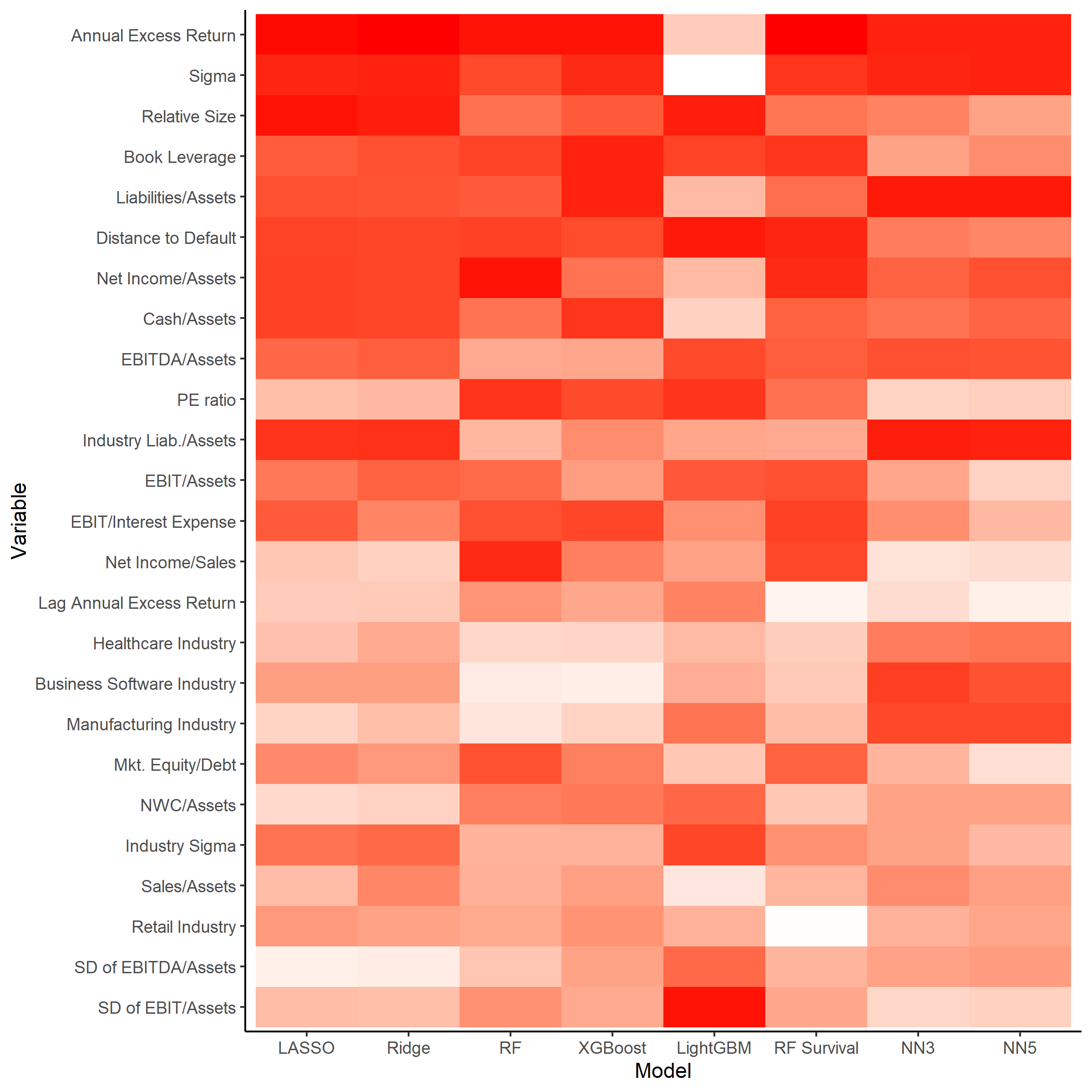}
	\label{fig:vimpDC}
\end{figure}

\begin{figure}[h!]
	\centering
	
	\caption{Variable Importance: Global Financial Crisis Years.}
	\includegraphics[width=1.1\textwidth]{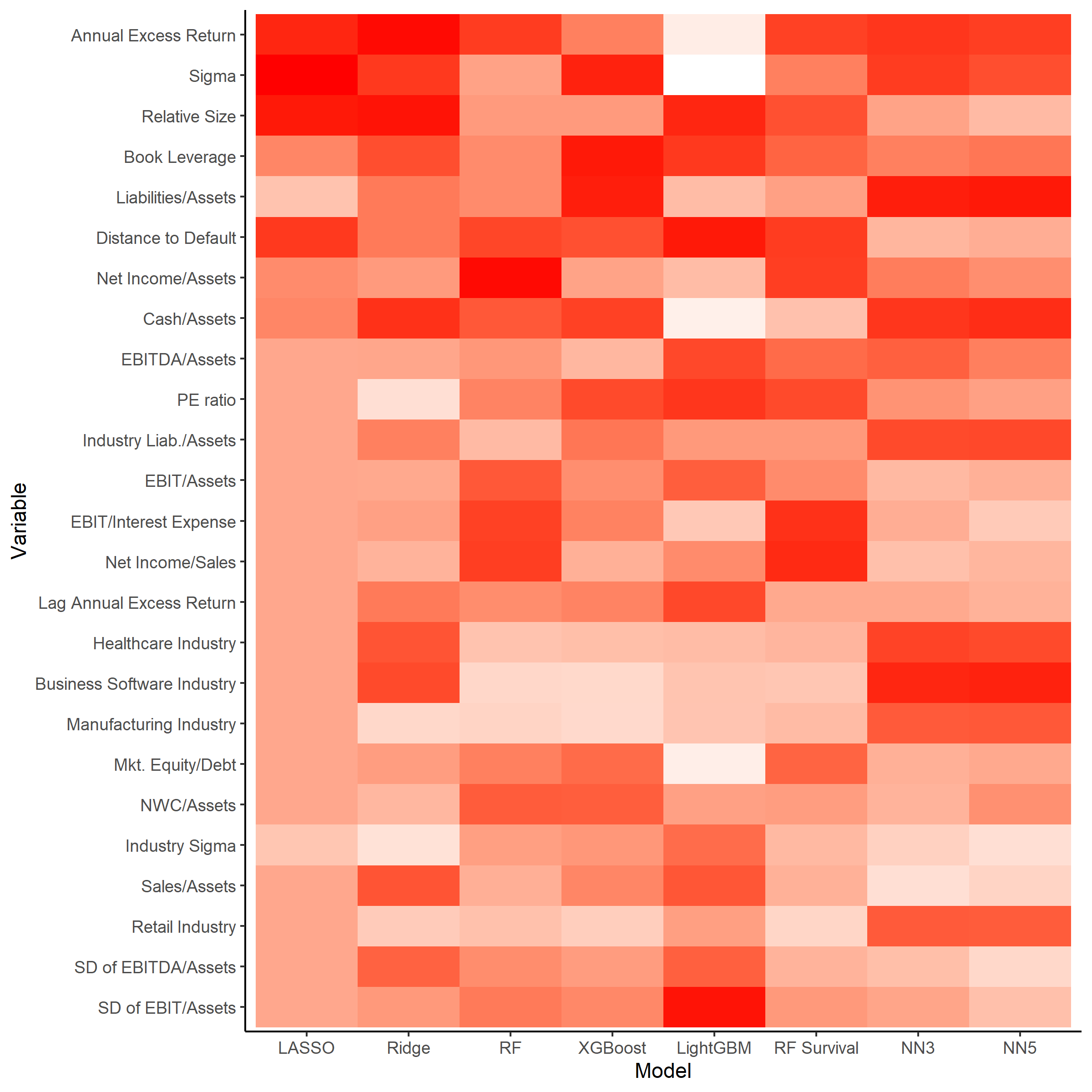}
	\label{fig:vimpGFC}
\end{figure}

Figures \ref{fig:vimpDC} and \ref{fig:vimpGFC} show the variable importance 
heatmaps for the dot-com and global financial crisis, respectively. 
Interestingly, for these crisis periods more predictors seem to matter and 
there is less agreement between models; that is, the dark red is more dispersed 
in the plot. 

In contrast with the full sample (Figure \ref{fig:vimp}), more accounting 
variables are considered important. For the dot-com years all models seem to 
agree that measures of liabilities---such as \emph{Book Leverage} (the ratio of 
long term debt to assets) and \emph{Liabilities/Assets}---are very relevant; 
other accounting measures coming from the Balance Sheet, such as \emph{Net 
Income/Assets} and \emph{Cash/Assets} seem to be more important. As well, all 
models agree that the \emph{Distance to Default} is significant.

Figure \ref{fig:vimpGFC} for the global financial crisis period shows 
interesting patterns. Outside of the LASSO and Ridge regressions, the market 
driven variables do not seem to be the most important ones and, again, 
accounting measures of firm liabilities are relevant for model performance.

Evidence from the variable importance measures suggests crisis periods are 
different. Though market variables are forward looking by nature and considered 
more timely than accounting information, financial statement ratios about the 
debt load of a firm seems to have predictive power over and above market 
variables during difficult economic conditions.

\begin{figure}[h!]
	\centering
	
	\caption{Variable Importance: Non-Crisis Years.}
	\includegraphics[width=1.1\textwidth]{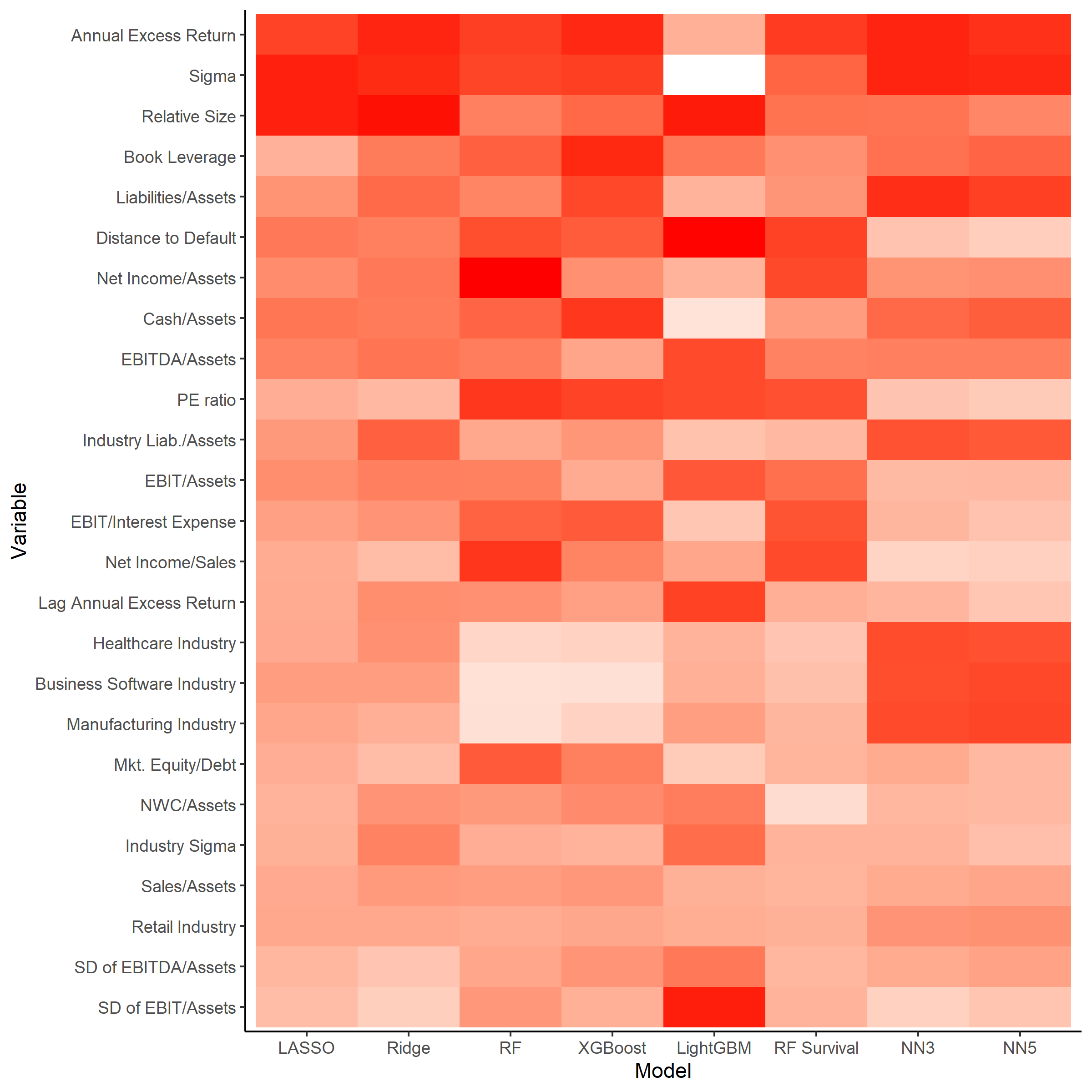}
	\label{fig:vimpNC}
\end{figure}

For completeness we evaluate the forecast performance for all years outside of 
the dot-com and global financial crisis. We present the AUC for these forecasts 
in Panel D of Table \ref{table:auc}, the boosted tree algorithms of XG Boost 
and LightGBM are the best performers with an AUC of 0.92. Figure 
\ref{fig:vimpNC} shows the variable importance heatmap for these years. Most 
methods agree that the market variables are the ones driving the forecasting 
power of the models, though again LightGBM assigns importance to the 
\emph{Distance to Default} and the standard deviation of \emph{EBIT/Assets}.

\subsection{Text Measures as Predictors}
In this section we test if inclusion of predictors based on text analysis of 10-K filings improve the performance of the model. Electronic filings on the SEC-EDGAR website start in 1993, thus our sample in these tests is smaller. We begin our forecast in the year 2000 and walk it forward one year at a time until 2019, we use the prior years beginning in 1993 for training and validation as described above.

We include seven total text measures: the four word count measures of \cite{loughran2011liability}, the Gunning-Fog index for readability, and the VADER and FinBERT sentiment measures. We start with a baseline model that includes as predictors all firm, industry and macroeconomic variables, then we add text measures one at a time.

Table \ref{table:txt} presents results for the out-of-sample AUC. The \emph{Baseline} column does not include any of the text measures as predictors, we include it because this is a different sample period from Table \ref{table:auc}, with shorter training and testing samples. The tree ensemble models exhibit the best performance in this sample, the survival random forest (RF Survival) has a particularly high AUC of 0.97 and the XGBoost model also exhibits better performance in this shorter time span.

The rest of the columns of Table \ref{table:txt} show the models' performance when the 10-K derived variables are included as predictors one at a time. None of the text measures seem to provide much benefit; the performance of most models does not improve in a meaningful way with the inclusion of any of the text measures. Some algorithms exhibit some improvement in AUC, such as the LightGBM which increases from 0.91 in the baseline to 0.93 with all text variables included, but this improvement is marginal at best. 

We believe our finding that these seven text measures do not improve model performance may not imply that unstructured or textual data is not useful for default prediction. It seems plausible that common lexicons used in the literature to gauge sentiment may not identify an event as rare and extreme as bankruptcy. Further, the changing nature of 10-K filings through the years represent a challenge for out-of-sample tests.

\subsection{Feature Reduction}
Several of our predictor variables, or features, have high pairwise correlation, for example, \emph{Book leverage} and the ratio of \emph{Liabilities/Assets} have correlation above 0.8. It is possible that some of our feature contribute little information to our prediction and instead add estimation error and complexity. In this section we investigate if reducing the number of predictors improves the performance of our models.

We reduce the number of predictors in two different ways. First, we use a small subset of variables that prior literature has shown to have forecasting power over default. In a different agnostic approach, we reduce the number of features through a principal component (PC) analysis and use a subset of PCs as predictors. Here we start with all of our predictor variables, including text measures, thus our out-of-sample period is from 2000 to 2019.

In the first approach, we use as predictors the variables from the market model of \cite{shumway2001forecasting} (\emph{Net Income/Assets, Liabilities/Assets, Annual Excess Return, Beta, Sigma} and \emph{Relative Size}) in addition to the \emph{Distance to Default} and the \emph{FinBERT} text measure. Row (1) of Table \ref{table:red} shows the the AUC for each algorithm with out-of-sample period from 2000 to 2019. Most of the algorithms have good performance with AUC above 0.90 except for the LASSO whose performance is hardly better than a random flip of the coin. The RF Survival is the best model with an AUC of 0.97, which seems remarkable given that the model has only 8 predictor variables.

In our second approach to feature reduction, we use principal components to reduce our parameter space. We select the number of PCs that explain 95\% of the variance of our predictors in the training data. Since we walk forward our training data one year at a time, we select PCs for every year we make a forecast. The benefit of this approach is that we are agnostic about the number of predictors and instead let the data speak for itself. For most forecast years, the number of PCs that explain 95\% of the variation in predictors is between 20 and 30.

Row (2) of Table \ref{table:red} shows the results from using PCs to reduce the number of predictors. Aside from the LASSO, all algorithms show a slight reduction in AUC in this approach relative to selecting variables based on prior literature. Nevertheless, the survival random forest remains the best model with an AUC of 0.97 for this forecasting period from 2000 to 2019. 

Results in this section show that it is possible to obtain very good forecasts with a smaller number of predictors, particularly when these predictors are chosen based on findings of prior literature.

\subsection{Economic Significance}
Our analysis so far assumes an equal cost to misclassification; that is, it is equally costly to predict a firm will default when it does not default, as it is to predict a firm will not default when it actually does. From a lender's perspective, it seems the latter situation is much costlier than the former. The losses from lending to a firm that defaults can be substantial, whereas the losses from not lending to a firm that does not default might just be an opportunity cost. 

In this section we extend our analysis by considering the differential cost of misclassification. We assess the economic impact of different prediction algorithms by following the competitive credit market model of \cite{agarwal2008comparing}. Each algorithm is a ``bank" than competes to fund a ``loan" of one year, where each company is a loan of equal size. As \cite{agarwal2008comparing} we normalize the loan market to \$100 million annually and assume each algorithm quotes a credit spread for each company as
\begin{equation}
    Spread = \frac{p(Y=1|X)}{p(Y=0|X)} LGD + k
    \label{eq:spread}
\end{equation}
where $p(Y=1|X)$ is the conditional probability of default computed by the algorithm, $LGD$ is the loss given default and $k$ is the spread for the highest quality borrower. We assume the loss given default is 45\% and use the spread paid by Aaa borrowers at the close of the prior year as measure of $k$. 

Each algorithm quotes a spread for every company based on equation \eqref{eq:spread} and the algorithm with the lowest spread is the one that funds the loan. 

Table \ref{table:econ} presents the economic gains and losses from following this simple model of credit competition of \cite{agarwal2008comparing} between algorithms. In panel A, we use the out-of-sample predictions from our baseline model whose results were presented in the last column of Table \ref{table:auc}, the sample is from 1990 to 2019. In panel B we use the more parsimonious model with reduced predictors presented in row (1) of Table \ref{table:red}, in this panel the sample is reduced to 2000 to 2019. Equation \eqref{eq:spread} is used to compute the credit spreads for each algorithm. The table presents the number of loans funded, how many of these loans ended in default, and the market share---measured as the percentage of total loans funded---for each algorithm. 

In this simulated credit market, if an algorithm assigns the lowest default probability among the algorithms to a loan that does not default, it increases its economic profit by capturing the earned interest income on the loan; but if it assigns the lowest probability among algorithms to a loan that does default then it suffers a large economic loss due to the lost principal amount. We observe that the survival random forest (RF Survival) and the 5 layer neural network (NN5) fund the majority of the loans in both panels, implying these models assigned the lowest default probability in most cases. The gradient boosted tree algorithms, XG Boost and LightGBM, which had the best AUC in the overall sample, fund less than 6\% of all loans in panel A and about 18\% in panel B. The RF Survival funds over half of all loans and its default rate, while above the median for all the algorithms, does not seem to be an outlier.

The last four columns of the table show gains and losses for each algorithm. The \emph{Annual Interest Income} is the average annual interest generated by each loan funded that did not default, while the \emph{Annual Losses} is the average annual loss due to loans funded that defaulted, the \emph{Annual Profit} is the difference between the interest income and losses, and the \emph{ROA} is the total profit divided by the face value of the loans funded. 

In both panels, the annual profit of the survival random forest is much higher than that of competing models, which is unsurprising given its very large market share in the simulation and its elevated but not outlying default rate; on the other hand, the ROA of the survival random forest is the third lowest (lowest) in panel A (B). It is not clear a priori which would be the preferred metric for a lender employing these predictive models, however it is worth noting that models with the higher ROA---such as the LASSO and Ridge---seem to be extremely selective in the loans they fund and thus have minuscule market shares.

Once we allow for an asymmetric cost to misclassification, the findings in Table \ref{table:econ} suggest that the survival random forest algorithm can be a competitive model in practice since it captures large dollar profits. It is interesting to note that the pattern of the results is similar in both panels, though panel B has only a handful of predictors that were chosen based on previous literature.

\section{Conclusion}

In this paper we evaluate the ability of different machine learning algorithms to forecast corporate bankruptcy. We use an extensive sample of bankruptcies from 1969 and conduct one-year-ahead out-of-sample forecasts for 1990 to 2019. We evaluate model performance by computing the receiver operating characteristic area under the curve (AUC) for the predictions. Boosted trees algorithms (XG Boost and LightGBM) perform best, reaching an AUC of 0.92.

Our findings suggest most of the performance comes from firm level variables, industry, and macroeconomic indicators do not seem to improve forecasts. For our overall sample, stock market variables---such as a firm's annual return, relative market size, and unsystematic risk---drive forecast performance. When we look at two notable crises in our out-of-sample period, the burst of the dot-com bubble and the global financial crisis, variable importance is more mixed. We find that financial statement information, in particular accounting ratios that reflect a firm's liabilities load, become more relevant for forecasting during the crisis. 

Interestingly, even during crisis periods industry and macroeconomic variables do not seem to improve forecast performance. This finding suggests default prediction remains mostly a firm-level event; though it is likely that default losses are impacted by industry- and economy-wide states \citep{chava2011modeling}.

We use the simulated loan competition model of \cite{agarwal2008comparing} to gauge the economic significance of our predictions. Our findings suggest that ensemble tree models are well suited for practice, in particular the survival random forest algorithm of \cite{ishwaran2008random} exhibits large dollar profits.

The predictive accuracy of our ML models does not improve in a meaningful way when we include features based on companies' annual 10-K filings. We believe bankruptcy, as a rare and extreme event, may present a particular challenge for unstructured data.

\singlespacing
\newpage
\bibliographystyle{apsr}
\bibliography{references}

\newpage

\newpage

\newpage

\clearpage
\begin{figure}[h!]
	\caption{Network graph of an \textit{L} hidden layer NN.}
	\vspace{12pt}
	\include{neuralnet}
	\label{fig:net}
\end{figure}

\newpage

\newpage

\newpage

\newpage

\newpage

\newpage
\newpage

\include{summary}

\newpage

\include{auc}

\newpage

\include{text}

\newpage

\include{var_reduction}

\newpage

\include{econ_sign}

\end{document}

%% file: neuralnet.tex
%
%
\usetikzlibrary{fadings}

%

		\centering
		\begin{tikzpicture}[shorten >=1pt]
			\tikzstyle{unit}=[draw,shape=circle,minimum size=1.15cm]
			\tikzstyle{hidden}=[draw,shape=circle,minimum size=1.15cm]
			
			\node[unit](x0) at (0,4){$x_0$};
			\node[unit](x1) at (0,2.5){$x_1$};
			\node at (0,1.1){\vdots};
			\node[unit](xd) at (0,-0.5){$x_D$};
			
			\node[hidden](h10) at (3,4){$y_0^{(1)}$};
			\node[hidden](h11) at (3,2.5){$y_1^{(1)}$};
			\node at (3,1.1){\vdots};
			\node[hidden](h1m) at (3,-0.5){$y_{m^{(1)}}^{(1)}$};
			
			\node(h22) at (5,0){};
			\node(h21) at (5,2){};
			\node(h20) at (5,4){};
			
			\node(d3) at (6,0){$\ldots$};
			\node(d2) at (6,2){$\ldots$};
			\node(d1) at (6,4){$\ldots$};
			
			\node(hL12) at (7,0){};
			\node(hL11) at (7,2){};
			\node(hL10) at (7,4){};
			
			\node[hidden](hL0) at (9,4){$y_0^{(L)}$};
			\node[hidden](hL1) at (9,2.5){$y_1^{(L)}$};
			\node at (9,1.1){\vdots};
			\node[hidden](hLm) at (9,-0.5){$y_{m^{(L)}}^{(L)}$};
			
			\node[unit](y1) at (12,2.5){$y_1^{(L+1)}$};
			
			\draw[->] (x0) -- (h11);
			\draw[->] (x0) -- (h1m);
			
			\draw[->] (x1) -- (h11);
			\draw[->] (x1) -- (h1m);
			
			\draw[->] (xd) -- (h11);
			\draw[->] (xd) -- (h1m);
			
			\draw[->] (hL0) -- (y1);
			
			\draw[->] (hL1) -- (y1);
			
			\draw[->] (hLm) -- (y1);
			
			\draw[->,path fading=east] (h10) -- (h21);
			\draw[->,path fading=east] (h10) -- (h22);
			
			\draw[->,path fading=east] (h11) -- (h21);
			\draw[->,path fading=east] (h11) -- (h22);
			
			\draw[->,path fading=east] (h1m) -- (h21);
			\draw[->,path fading=east] (h1m) -- (h22);
			
			\draw[->,path fading=west] (hL10) -- (hL1);
			\draw[->,path fading=west] (hL11) -- (hL1);
			\draw[->,path fading=west] (hL12) -- (hL1);
			
			\draw[->,path fading=west] (hL10) -- (hLm);
			\draw[->,path fading=west] (hL11) -- (hLm);
			\draw[->,path fading=west] (hL12) -- (hLm);
			
			\draw 
			[decorate,decoration={brace,amplitude=10pt},xshift=-4pt,yshift=0pt] 
			(-0.5,5) -- (0.75,5) node [black,midway,yshift=+0.6cm]{input 
			layer};
			\draw 
			[decorate,decoration={brace,amplitude=10pt},xshift=-4pt,yshift=0pt] 
			(2.5,5) -- (3.75,5) node 
			[black,midway,yshift=+0.6cm]{$1^{\text{st}}$ hidden layer};
			\draw 
			[decorate,decoration={brace,amplitude=10pt},xshift=-4pt,yshift=0pt] 
			(8.5,5) -- (9.75,5) node 
			[black,midway,yshift=+0.6cm]{$L^{\text{th}}$ hidden layer};
			\draw 
			[decorate,decoration={brace,amplitude=10pt},xshift=-4pt,yshift=0pt] 
			(11.5,5) -- (12.75,5) node [black,midway,yshift=+0.6cm]{output 
			layer};
		\end{tikzpicture}


%% file: timeline.tex
%

\definecolor{myLightGray}{RGB}{191,191,191}
\definecolor{myGray}{RGB}{160,160,160}
\definecolor{myDarkGray}{RGB}{144,144,144}
\definecolor{myRed}{RGB}{255,58,70}
\definecolor{myGreen}{RGB}{0,255,71}

%
	
\begin{center}
	
\begin{subfigure}[t]{\textwidth}
	\resizebox{\textwidth}{!}{
	\begin{tikzpicture}[%
		every node/.style={
			font=\scriptsize,
			text height=1ex,
			text depth=.25ex,
		},
		]
		\draw[<->] (0,0) -- (8,0);
		
		\foreach \x in {0.75,1.5,3,4,5,6,7}{
			\draw (\x cm,3pt) -- (\x cm,0pt);
		}
		
		\node[anchor=north] at (0.75,0) {$1969$};
		\node[anchor=north] at (1.5,0) {$1970$};
		\node[anchor=north] at (2.2,0) {$\cdots$};
		\node[anchor=north] at (3,0) {$t-3$};
		\node[anchor=north] at (4,0) {$t-2$};
		\node[anchor=north] at (5,0) {$t-1$};
		\node[anchor=north] at (6,0) {$t$};
		\node[anchor=north] at (7,0) {$t+1$};
		\node[anchor=north] at (8,0) {years};
		
		\fill[myRed, opacity=0.8] (0.75,0) rectangle (3,0.25);	
		
		\fill[myRed, opacity=0.4] (3,0) rectangle (5,0.25);	
		
		\fill[myGreen, opacity=0.5] (5,0) rectangle (6,0.25);
		
		\draw[decorate,decoration={brace,amplitude=5pt}] (3.3,-0.4) -- 
		(0.7,-0.4) 
		node[anchor=north,midway,below=4pt] {Training};
		
		\draw[decorate,decoration={brace,amplitude=5pt}] (5.2,-0.4) -- 
		(3.55,-0.4) 
		node[anchor=north,midway,below=4pt] {Validation};
		
		\draw[decorate,decoration={brace,amplitude=5pt}] (6.25,-0.4) -- 
		(5.75,-0.4)
		node[anchor=north,midway,below=4pt] {Testing};
	\end{tikzpicture}
}
\caption{Forecast for year $t$}
\end{subfigure}

\vspace{1in}

\begin{subfigure}[t]{\textwidth}
	\resizebox{\textwidth}{!}{
		\begin{tikzpicture}[%
			every node/.style={
				font=\scriptsize,
				text height=1ex,
				text depth=.25ex,
			},
			]
			\draw[<->] (0,0) -- (8,0);
			
			\foreach \x in {0.75,1.5,3,4,5,6,7}{
				\draw (\x cm,3pt) -- (\x cm,0pt);
			}
			
			\node[anchor=north] at (0.75,0) {$1969$};
			\node[anchor=north] at (1.5,0) {$1970$};
			\node[anchor=north] at (2.2,0) {$\cdots$};
			\node[anchor=north] at (3,0) {$t-3$};
			\node[anchor=north] at (4,0) {$t-2$};
			\node[anchor=north] at (5,0) {$t-1$};
			\node[anchor=north] at (6,0) {$t$};
			\node[anchor=north] at (7,0) {$t+1$};
			\node[anchor=north] at (8,0) {years};
			
			\fill[myRed, opacity=0.8] (0.75,0) rectangle (4,0.25);	
			
			\fill[myRed, opacity=0.4] (4,0) rectangle (6,0.25);	
			
			\fill[myGreen, opacity=0.5] (6,0) rectangle (7,0.25);
			
			\draw[decorate,decoration={brace,amplitude=5pt}] (4.3,-0.4) -- 
			(0.7,-0.4) 
			node[anchor=north,midway,below=4pt] {Training};
			
			\draw[decorate,decoration={brace,amplitude=5pt}] (6.2,-0.4) -- 
			(4.55,-0.4) 
			node[anchor=north,midway,below=4pt] {Validation};
			
			\draw[decorate,decoration={brace,amplitude=5pt}] (7.25,-0.4) -- 
			(6.75,-0.4)
			node[anchor=north,midway,below=4pt] {Testing};
		\end{tikzpicture}
	}
	\caption{Forecast for year $t+1$}
\end{subfigure}
\end{center}

%

%% file: summary.tex
\begin{table}[t!]  
  \caption{\textbf{Summary Statistics}. } \vspace{-1mm} 
 \begin{flushleft} This table presents summary
                                      statistics for the predictor variables used in our study. The sample
                                      comprises 131,261 firm-year observations from 1969 to 2019 with 2,585 instances of default. For the text-based variables the sample is from 1993 to 2019.  \vspace{-5mm} 
 \end{flushleft} 
  \label{table:summary} 
\resizebox{\textwidth}{!} { 

                        \begin{tabular*}{1.2\textwidth}{@{\extracolsep{\fill}}l*{5}{c}}
\\[-1.8ex]\hline 
\hline \\[-1.8ex] 
 & \multicolumn{1}{c}{Mean} & \multicolumn{1}{c}{St. Dev.} & \multicolumn{1}{c}{Pctl(25)} & \multicolumn{1}{c}{Median} & \multicolumn{1}{c}{Pctl(75)} \\ 
\hline \\[-1.8ex] 
Distance to Default & 5.996 & 35.448 & 1.875 & 4.168 & 7.007 \\ 
[1ex] \textit{Accounting Variables} \\ 
 Net Income/Assets & $-$0.027 & 0.789 & $-$0.001 & 0.026 & 0.057 \\ 
Liabilities/Assets & 0.650 & 0.553 & 0.489 & 0.616 & 0.777 \\ 
NWC/Assets & 0.143 & 0.561 & 0.014 & 0.146 & 0.302 \\ 
Retained Earnings/Assets & $-$0.157 & 3.939 & $-$0.028 & 0.099 & 0.278 \\ 
EBIT/Assets & 0.031 & 0.527 & 0.019 & 0.064 & 0.109 \\ 
Sales/Assets & 1.060 & 0.965 & 0.345 & 0.930 & 1.489 \\ 
Net Income/Equity & $-$0.413 & 109.326 & 0.007 & 0.091 & 0.148 \\ 
Net Income/Sales & $-$0.781 & 73.454 & $-$0.001 & 0.036 & 0.083 \\ 
SD of NI/Sales & 1.040 & 35.607 & 0.013 & 0.031 & 0.084 \\ 
SD of NI/Equity & 1.318 & 77.168 & 0.026 & 0.065 & 0.190 \\ 
SD of EBIT/Assets & 0.059 & 0.330 & 0.013 & 0.028 & 0.057 \\ 
Book Leverage & 0.347 & 0.309 & 0.196 & 0.308 & 0.446 \\ 
EBIT/Interest Expense & 9.414 & 124.645 & 1.048 & 3.047 & 7.764 \\ 
Operating Cost/Assets & 0.985 & 1.089 & 0.269 & 0.828 & 1.382 \\ 
Cash/Assets & 0.059 & 0.074 & 0.015 & 0.036 & 0.075 \\ 
Gross Profit/Sales & $-$0.285 & 70.410 & 0.207 & 0.315 & 0.467 \\ 
Net PPE/Assets & 0.319 & 0.265 & 0.088 & 0.264 & 0.501 \\ 
EBITDA/Assets & 0.074 & 0.509 & 0.034 & 0.102 & 0.155 \\ 
SD of EBITDA/Assets & 0.056 & 0.315 & 0.013 & 0.028 & 0.056 \\ 
[1ex] \textit{Market Variables} \\ 
 Mkt. Equity/Debt & 4.989 & 113.136 & 0.765 & 1.693 & 3.830 \\ 
Annual Excess Return & 0.006 & 0.714 & $-$0.316 & $-$0.061 & 0.196 \\ 
Beta & 1.065 & 1.567 & 0.333 & 0.958 & 1.662 \\ 
Sigma & 0.118 & 0.099 & 0.062 & 0.093 & 0.144 \\ 
Relative Size & $-$3.636 & 2.180 & $-$5.208 & $-$3.658 & $-$2.116 \\ 
PE ratio & 17.469 & 541.966 & $-$0.113 & 10.865 & 18.650 \\ 
Lag Annual Excess Return & 0.016 & 0.670 & $-$0.307 & $-$0.059 & 0.198 \\ 
[1ex] \textit{Industry Variables} \\ 
 HH Index of Sales & 0.039 & 0.043 & 0.017 & 0.025 & 0.039 \\ 
Industry Liab./Assets & 0.579 & 0.155 & 0.491 & 0.539 & 0.622 \\ 
Industry Sigma & 0.101 & 0.033 & 0.078 & 0.095 & 0.120 \\ 
Industry Defaults Last 1 Yr & 6.684 & 7.264 & 1.000 & 4.000 & 10.000 \\ 
Industry Defaults Last 2 Yrs & 13.130 & 13.447 & 3.000 & 8.000 & 19.000 \\ 
\end{tabular*} 
 }
\end{table} 

\clearpage

\begin{table}[t]  
  \ContinuedFloat 
 \caption{\textbf{Summary Statistics} (continued) } \vspace{2mm} 
 \begin{flushleft}  \vspace{2mm} 
 \end{flushleft} 
 \resizebox{\textwidth}{!} { 

                        \begin{tabular*}{1.2\textwidth}{@{\extracolsep{\fill}}l*{5}{c}}
\\[-1.8ex]\hline 
\hline \\[-1.8ex] 
 & \multicolumn{1}{c}{Mean} & \multicolumn{1}{c}{St. Dev.} & \multicolumn{1}{c}{Pctl(25)} & \multicolumn{1}{c}{Median} & \multicolumn{1}{c}{Pctl(75)} \\ 
\hline \\[-1.8ex] 
\textit{Macroeconomic Variables} \\ 
 Term Spread & 1.134 & 1.163 & 0.170 & 1.030 & 1.930 \\ 
Credit Spread & 1.100 & 0.486 & 0.780 & 0.980 & 1.270 \\ 
Recession Dummy & 0.116 & 0.320 & 0.000 & 0.000 & 0.000 \\ 
Inflation & 3.648 & 2.585 & 2.035 & 2.981 & 4.332 \\
GDP Growth & 2.775 & 3.138 & 1.114 & 3.007 & 4.661 \\
Industrial Production & 2.386 & 4.229 & 0.930 & 2.611 & 4.980 \\
Unemployment Rate & 6.029 & 1.515 & 5.000 & 5.700 & 7.200 \\
[1ex] \textit{Text-Based Variables} \\ 
Positive Words & 0.064 & 0.133 & 0.000 & 0.026 & 0.077 \\ 
Negative Words & 0.121 & 0.217 & 0.000 & 0.059 & 0.142 \\ 
Uncertainty Words & 0.100 & 0.198 & 0.000 & 0.048 & 0.116 \\ 
Litigious Words & 0.096 & 0.265 & 0.000 & 0.035 & 0.104 \\ 
Gunning-Fog Index & 10.050 & 1.130 & 9.290 & 9.940 & 10.680 \\ 
VADER Sentiment & 0.221 & 0.050 & 0.192 & 0.223 & 0.252 \\ 
FinBERT Sentiment & 0.008 & 0.045 & $-$0.020 & 0.009 & 0.037 \\ 
\hline \\[-1.8ex] 
\end{tabular*} 
 }
\end{table}

%% file: auc.tex
\begin{table}[!htbp]  
  \caption{\textbf{Model Performance}. } \vspace{2mm} 
 \begin{flushleft} This table presents the 
          out-of-sample performance for all algorithms based on the Area Under the Curve
          (AUC) measure. The column $DD$ only uses the distance to default and 
          the Fama-French 
          12 industry dummies as predictors. Subsequent columns add the group of variables indicated 
          at the top to the group of predictors.
          Panel A shows the performance for our full out-of-sample predictions 
          from 1990 to 2019. Panel B shows model performance during the dot-com 
          bubble years
          of 1999 to 2001. Panel C shows performance during the Global Financial Crisis of
          2007 to 2009. Panel D shows the performance for all years outside of the two crisis
          periods.  \vspace{2mm} 
 \end{flushleft}

          \emph{Panel A: Full Sample, 1990--2019} 
  
  \label{table:auc} 
\resizebox{\textwidth}{!} { 

                        \begin{tabular*}{1\textwidth}{@{\extracolsep{\fill}}l*{5}{c}}
\\[-1.8ex]\hline 
\hline \\[-1.8ex] 
Algorithm & DD & + Accounting & + Market & + Industry & + Macro \\ 
\hline \\[-1.8ex] 
LASSO & $0.551$ & $0.646$ & $0.841$ & $0.841$ & $0.852$ \\ 
Ridge & $0.563$ & $0.574$ & $0.812$ & $0.860$ & $0.847$ \\ 
Random Forest & $0.833$ & $0.896$ & $0.906$ & $0.908$ & $0.909$ \\ 
XG Boost & $0.837$ & $0.907$ & $0.916$ & $0.918$ & $0.915$ \\ 
LightGBM & $0.832$ & $0.903$ & $0.913$ & $0.914$ & $0.915$ \\ 
RF Survival & $0.831$ & $0.831$ & $0.865$ & $0.859$ & $0.872$ \\ 
NN3 & $0.779$ & $0.853$ & $0.894$ & $0.891$ & $0.890$ \\ 
NN5 & $0.807$ & $0.864$ & $0.897$ & $0.890$ & $0.886$ \\ 
\hline \\[-1.8ex] 
\end{tabular*} 
 }

 \begin{flushleft} \vspace{20pt}  \vspace{2mm} 
 \end{flushleft}

          \emph{Panel B: Dot-com Bubble, 1999--2001} 
  
  \label{} 
\resizebox{\textwidth}{!} { 

                        \begin{tabular*}{1\textwidth}{@{\extracolsep{\fill}}l*{5}{c}}
\\[-1.8ex]\hline 
\hline \\[-1.8ex] 
Algorithm & DD & + Accounting & + Market & + Industry & + Macro \\ 
\hline \\[-1.8ex] 
LASSO & $0.570$ & $0.802$ & $0.838$ & $0.839$ & $0.839$ \\ 
Ridge & $0.577$ & $0.717$ & $0.799$ & $0.835$ & $0.771$ \\ 
Random Forest & $0.820$ & $0.890$ & $0.898$ & $0.899$ & $0.899$ \\ 
XG Boost & $0.819$ & $0.898$ & $0.909$ & $0.907$ & $0.908$ \\ 
LightGBM & $0.815$ & $0.890$ & $0.901$ & $0.905$ & $0.901$ \\ 
RF Survival & $0.880$ & $0.880$ & $0.871$ & $0.888$ & $0.890$ \\ 
NN3 & $0.740$ & $0.828$ & $0.870$ & $0.868$ & $0.869$ \\ 
NN5 & $0.781$ & $0.866$ & $0.869$ & $0.861$ & $0.861$ \\ 
\hline \\[-1.8ex] 
\end{tabular*} 
 }
\end{table} 

\begin{table}[!htbp]  
  \ContinuedFloat 
 \caption{\textbf{Model Performance.} (continued) 
                } \vspace{2mm} 
 \begin{flushleft}  \vspace{2mm} 
 \end{flushleft}

          \emph{Panel C: Global Financial Crisis, 2007--2009} 
  
  \label{} 
\resizebox{\textwidth}{!} { 

                        \begin{tabular*}{1\textwidth}{@{\extracolsep{\fill}}l*{5}{c}}
\\[-1.8ex]\hline 
\hline \\[-1.8ex] 
Algorithm & DD & + Accounting & + Market & + Industry & + Macro \\ 
\hline \\[-1.8ex] 
LASSO & $0.529$ & $0.705$ & $0.887$ & $0.884$ & $0.884$ \\ 
Ridge & $0.547$ & $0.700$ & $0.869$ & $0.888$ & $0.882$ \\ 
Random Forest & $0.799$ & $0.900$ & $0.909$ & $0.910$ & $0.908$ \\ 
XG Boost & $0.802$ & $0.898$ & $0.911$ & $0.912$ & $0.913$ \\ 
LightGBM & $0.796$ & $0.890$ & $0.900$ & $0.906$ & $0.916$ \\ 
RF Survival & $0.850$ & $0.850$ & $0.885$ & $0.881$ & $0.893$ \\ 
NN3 & $0.717$ & $0.876$ & $0.906$ & $0.907$ & $0.910$ \\ 
NN5 & $0.761$ & $0.868$ & $0.912$ & $0.914$ & $0.898$ \\ 
\hline \\[-1.8ex] 
\end{tabular*} 
 }

 \begin{flushleft} \vspace{20pt}  \vspace{2mm} 
 \end{flushleft}

          \emph{Panel D: Non-crisis periods} 
  
  \label{} 
\resizebox{\textwidth}{!} { 

                        \begin{tabular*}{1\textwidth}{@{\extracolsep{\fill}}l*{5}{c}}
\\[-1.8ex]\hline 
\hline \\[-1.8ex] 
Algorithm & DD & + Accounting & + Market & + Industry & + Macro \\ 
\hline \\[-1.8ex] 
LASSO & $0.551$ & $0.620$ & $0.828$ & $0.828$ & $0.828$ \\ 
Ridge & $0.560$ & $0.583$ & $0.837$ & $0.858$ & $0.858$ \\ 
Random Forest & $0.832$ & $0.894$ & $0.905$ & $0.907$ & $0.908$ \\ 
XG Boost & $0.838$ & $0.908$ & $0.915$ & $0.919$ & $0.915$ \\ 
LightGBM & $0.834$ & $0.907$ & $0.916$ & $0.914$ & $0.915$ \\ 
RF Survival & $0.807$ & $0.807$ & $0.854$ & $0.840$ & $0.855$ \\ 
NN3 & $0.788$ & $0.855$ & $0.896$ & $0.894$ & $0.892$ \\ 
NN5 & $0.810$ & $0.867$ & $0.901$ & $0.893$ & $0.890$ \\ 
\hline \\[-1.8ex] 
\end{tabular*} 
 }
\end{table}

%% file: text.tex
\begin{sidewaystable}[!htbp]  
  \caption{\textbf{Text Measures}. } \vspace{2mm} 
 \begin{flushleft} This table presents the out-of-sample performance for all algorithms based on the Area Under the Curve (AUC) measure when we include text measures from 10-K filings as predictors. The training sample for all algorithms begins in 1993, the out-of-sample is from 2000 to 2019. The column \emph{Baseline} includes all non-text predictors, subsequent columns add the text measure indicated at the top to the group of predictors. The first four text measures (\emph{Positive, Negative, Uncertain, Litigious} are from \cite{loughran2011liability}, \emph{GFI} is the Gunning-Fog index, \emph{VADER} is the Valence Aware Dictionary and sEntiment Reasoner, and \emph{FinBERT} is the BERT model trained on financial data.  \vspace{2mm} 
 \end{flushleft}  
  \label{table:txt} 
\resizebox{\textwidth}{!} { 

                        \begin{tabular*}{1.15\textwidth}{@{\extracolsep{\fill}}l*{8}{c}}
\\[-1.8ex]\hline 
\hline \\[-1.8ex] 
Method & Baseline & + Positive & + Negative & + Uncertain & + Litigious & + GFI & + VADER & + FinBERT \\ 
\hline \\[-1.8ex] 
LASSO & $0.472$ & $0.472$ & $0.472$ & $0.472$ & $0.472$ & $0.472$ & $0.472$ & $0.472$ \\ 
Ridge & $0.883$ & $0.884$ & $0.887$ & $0.888$ & $0.888$ & $0.885$ & $0.883$ & $0.889$ \\ 
Random Forest & $0.923$ & $0.922$ & $0.925$ & $0.923$ & $0.923$ & $0.925$ & $0.926$ & $0.927$ \\ 
XG Boost & $0.929$ & $0.932$ & $0.917$ & $0.927$ & $0.926$ & $0.933$ & $0.934$ & $0.933$ \\ 
LightGBM & $0.913$ & $0.927$ & $0.913$ & $0.922$ & $0.926$ & $0.917$ & $0.926$ & $0.927$ \\ 
RF Survival & $0.969$ & $0.969$ & $0.968$ & $0.965$ & $0.966$ & $0.964$ & $0.964$ & $0.970$ \\ 
NN3 & $0.847$ & $0.824$ & $0.830$ & $0.819$ & $0.827$ & $0.820$ & $0.831$ & $0.824$ \\ 
NN5 & $0.833$ & $0.841$ & $0.833$ & $0.807$ & $0.837$ & $0.800$ & $0.783$ & $0.821$ \\ 
\hline \\[-1.8ex] 
\end{tabular*} 
 }
\end{sidewaystable}

%% file: var_reduction.tex
\begin{table}[!htbp]  
  \caption{\textbf{Variable Reduction}. } \vspace{2mm} 
 \begin{flushleft} This table evaluates the robustness of our results reducing the number of predictor variables. The table presents the 
          out-of-sample performance for all algorithms based on the Area Under the Curve
          (AUC) measure for the period from 2000 to 2019. In row (1) the predictors are Distance to Default, net income/assets, liabilities/assets, sigma, annual excess return, beta, relative size, and the text measure FinBERT. In row (2) the predictors are based on principal components where we select the number of pricipal components that explain 95\% of the predictor variance in the training sample.  \vspace{2mm} 
 \end{flushleft}  
  \label{table:red} 
\resizebox{\textwidth}{!} { 

                        \begin{tabular*}{1.05\textwidth}{@{\extracolsep{\fill}}l*{8}{c}}
\\[-1.8ex]\hline 
\hline \\[-1.8ex] 
 & LASSO & Ridge & Random Forest & XG Boost & LightGBM & RF Survival & NN3 & NN5 \\ 
\hline \\[-1.8ex] 
(1) & $0.539$ & $0.858$ & $0.910$ & $0.912$ & $0.907$ & $0.965$ & $0.901$ & $0.895$ \\ 
(2) & $0.796$ & $0.783$ & $0.876$ & $0.900$ & $0.900$ & $0.965$ & $0.850$ & $0.812$ \\ 
\hline \\[-1.8ex] 
\end{tabular*} 
 }
\end{table}

%% file: econ_sign.tex
\begin{sidewaystable}[!htbp]  
  \caption{\textbf{Economic Significance}. } \vspace{2mm} 
 \begin{flushleft} This table evaluates the economic significance of the default predictions. We use the model of \cite{agarwal2008comparing} to simulate a loan competition market. In this model each company represents a loan of equal size, where the total market size is normalized to \$100 million every year. Each algorithm competes to fund the loan and the one that offers the lower rate makes the loan. In Panel A, the default predictions are from the models in the last column of panel A in Table \ref{table:auc}, the out-of-sample period is from 1990 to 2019; in Panel B, the default predictions are from the models with a reduced number of predictors in row (1) of Table \ref{table:red}, the out-of-sample period is from 2000 to 2019.  \emph{Loans Funded} is the number of loans each algorithm makes, \emph{Loans Defaulted} is the number of loans which default, \emph{Market Share} is the percentage of total loans funded by each algorithm. \emph{Annual Interest Income} is the average annual interest collected (in thousands) from the loans funded, \emph{Annual Losses} is the average annual loss from the defaulted loans, \emph{Annual Profit} is the difference between interest income and losses, and \emph{ROA} is the ratio of annual profit to annual assets.  \vspace{1mm} 
 \end{flushleft}  
  \label{table:econ} 
\resizebox{\textwidth}{!} { 

                        \begin{tabular*}{1.05\textwidth}{@{\extracolsep{\fill}}l*{8}{c}}
\\[-1.8ex]\hline 
\hline \\[-1.8ex] 
Algorithm & \thead{Loans \\ Funded} & \thead{Loans \\ Defaulted} & \thead{Default \\ Rate (\%)} & \thead{Market \\ Share (\%)} & \thead{Annual Interest \\ Income} & \thead{Annual \\ Losses} & \thead{Annual \\ Profit} & ROA (\%) \\ 
\hline \\[-1.8ex] 
\multicolumn{9}{l}{\emph{Panel A: All variables and out-of-sample period of 1990-2019}} \\
LASSO & $2,143$ & $51$ & $2.38$ & $2.43$ & $171.12$ & $24.93$ & $146.19$ & $5.84$ \\ 
Ridge & $1,583$ & $28$ & $1.77$ & $1.80$ & $133.59$ & $14.27$ & $119.32$ & $6.25$ \\ 
Random Forest & $263$ & $6$ & $2.28$ & $0.30$ & $23.77$ & $4.03$ & $19.73$ & $4.66$ \\ 
XG Boost & $1,992$ & $44$ & $2.21$ & $2.26$ & $143.27$ & $20.99$ & $122.28$ & $5.44$ \\ 
Light GBM & $3,212$ & $65$ & $2.02$ & $3.65$ & $226.50$ & $29.95$ & $196.54$ & $5.68$ \\ 
RF Survival & $54,302$ & $1,275$ & $2.35$ & $61.62$ & $3,689.17$ & $606.04$ & $3,083.13$ & $5.07$ \\ 
NN3 & $6,521$ & $132$ & $2.02$ & $7.40$ & $460.73$ & $65.02$ & $395.71$ & $5.21$ \\ 
NN5 & $18,103$ & $396$ & $2.19$ & $20.54$ & $1,261.00$ & $199.16$ & $1,061.84$ & $5.03$ \\ 
\hline \\[0ex] 

\multicolumn{9}{l}{\emph{Panel B: Reduced variables and out-of-sample period of 2000-2019}} \\
LASSO & $694$ & $4$ & $0.58$ & $1.57$ & $83.61$ & $3.68$ & $79.93$ & $4.95$ \\ 
Ridge & $327$ & $4$ & $1.22$ & $0.74$ & $41.08$ & $3.92$ & $37.16$ & $4.90$ \\ 
Random Forest & $1,093$ & $7$ & $0.64$ & $2.48$ & $129.67$ & $5.74$ & $123.92$ & $4.97$ \\ 
XG Boost & $3,506$ & $36$ & $1.03$ & $7.95$ & $411.94$ & $33.12$ & $378.81$ & $4.72$ \\ 
Light GBM & $4,584$ & $49$ & $1.07$ & $10.40$ & $536.18$ & $45.82$ & $490.36$ & $4.77$ \\ 
RF Survival & $23,718$ & $271$ & $1.14$ & $53.80$ & $2,747.71$ & $257.82$ & $2,489.89$ & $4.63$ \\ 
NN3 & $4,470$ & $57$ & $1.28$ & $10.14$ & $532.29$ & $55.50$ & $476.79$ & $4.69$ \\ 
NN5 & $5,696$ & $62$ & $1.09$ & $12.92$ & $678.63$ & $58.88$ & $619.74$ & $4.82$ \\ 
\hline \\[-1.8ex]
\end{tabular*} 
 }
\end{sidewaystable}

%% file: references.bib
@article{chavapurnanandam2010rfs,
    author = {Chava, Sudheer and Purnanandam, Amiyatosh},
    title = "{Is Default Risk Negatively Related to Stock Returns?}",
    journal = {The Review of Financial Studies},
    volume = {23},
    number = {6},
    pages = {2523-2559},
    year = {2010},
    month = {01},
    issn = {0893-9454},
    doi = {10.1093/rfs/hhp107}
}

@article{chava2004bankruptcy,
	title={Bankruptcy prediction with industry effects},
	author={Chava, Sudheer and Jarrow, Robert A},
	journal={Review of Finance},
	volume={8},
	number={4},
	pages={537--569},
	year={2004},
	publisher={Oxford University Press}
}

@article{alanis2018shareholder,
	title={Shareholder bargaining power, debt overhang, and investment},
	author={Alanis, Emmanuel and Chava, Sudheer and Kumar, Praveen},
	journal={Review of Corporate Finance Studies},
	volume={7},
	number={2},
	pages={276--318},
	year={2018},
	publisher={Oxford University Press}
}

@article{barboza2017machine,
	title={Machine learning models and bankruptcy prediction},
	author={Barboza, Flavio and Kimura, Herbert and Altman, Edward},
	journal={Expert Systems with Applications},
	volume={83},
	pages={405--417},
	year={2017},
	publisher={Elsevier}
}

@article{shumway2001forecasting,
	title={Forecasting bankruptcy more accurately: A simple hazard model},
	author={Shumway, Tyler},
	journal={The Journal of Business},
	volume={74},
	number={1},
	pages={101--124},
	year={2001},
	publisher={JSTOR}
}

@article{altman1968financial,
	title={Financial ratios, discriminant analysis and the prediction of 
	corporate bankruptcy},
	author={Altman, Edward I},
	journal={The Journal of Finance},
	volume={23},
	number={4},
	pages={589--609},
	year={1968},
	publisher={JSTOR}
}

@book{taddy2019business,
	title={Business data science: Combining machine learning and economics to 
	optimize, automate, and accelerate business decisions},
	author={Taddy, Matt},
	year={2019},
	publisher={McGraw Hill Professional}
}

@book{breiman1984classification,
	title={Classification and regression trees},
	author={Breiman, Leo and Friedman, Jerome and Stone, Charles J and Olshen, 
	Richard A},
	year={1984},
	publisher={CRC press}
}

@article{breiman2001random,
	title={Random forests},
	author={Breiman, Leo},
	journal={Machine Learning},
	volume={45},
	number={1},
	pages={5--32},
	year={2001},
	publisher={Springer}
}

@article{rasekhschaffe2019machine,
	title={Machine learning for stock selection},
	author={Rasekhschaffe, Keywan Christian and Jones, Robert C},
	journal={Financial Analysts Journal},
	volume={75},
	number={3},
	pages={70--88},
	year={2019},
	publisher={Taylor \& Francis}
}

@inproceedings{chen2016xgboost,
	title={Xgboost: A scalable tree boosting system},
	author={Chen, Tianqi and Guestrin, Carlos},
	booktitle={Proceedings of the 22nd acm sigkdd international conference on 
	knowledge discovery and data mining},
	pages={785--794},
	year={2016}
}

@article{bharath2008,
	title={Forecasting default with the Merton distance to default model},
	author={Bharath, Sreedhar T and Shumway, Tyler},
	journal={The Review of Financial Studies},
	volume={21},
	number={3},
	pages={1339--1369},
	year={2008},
	publisher={Oxford University Press}
}

@article{merton1974,
	title={On the pricing of corporate debt: The risk structure of interest rates},
	author={Merton, Robert C},
	journal={The Journal of Finance},
	volume={29},
	number={2},
	pages={449--470},
	year={1974},
	publisher={Wiley Online Library}
}

@book{hastie2009elements,
	title={The elements of statistical learning: data mining, inference, and 
	prediction},
	author={Hastie, Trevor and Tibshirani, Robert and Friedman, Jerome},
	year={2009},
	publisher={Springer Science \& Business Media}
}

@book{aggarwal2018neural,
	title={Neural networks and deep learning},
	author={Aggarwal, Charu C},
	year={2018},
	publisher={Springer}
}

@article{ke2017lightgbm,
	title={Lightgbm: A highly efficient gradient boosting decision tree},
	author={Ke, Guolin and Meng, Qi and Finley, Thomas and Wang, Taifeng and 
	Chen, Wei and Ma, Weidong and Ye, Qiwei and Liu, Tie-Yan},
	journal={Advances in neural information processing systems},
	volume={30},
	pages={3146--3154},
	year={2017}
}

@article{gu2020empirical,
	title={Empirical asset pricing via machine learning},
	author={Gu, Shihao and Kelly, Bryan and Xiu, Dacheng},
	journal={The Review of Financial Studies},
	volume={33},
	number={5},
	pages={2223--2273},
	year={2020},
	publisher={Oxford University Press}
}

@article{ishwaran2008random,
	title={Random survival forests},
	author={Ishwaran, Hemant and Kogalur, Udaya B and Blackstone, Eugene H and 
	Lauer, Michael S},
	journal={The annals of applied statistics},
	volume={2},
	number={3},
	pages={841--860},
	year={2008},
	publisher={Institute of Mathematical Statistics}
}

@article{chava2011modeling,
	title={Modeling the loss distribution},
	author={Chava, Sudheer and Stefanescu, Catalina and Turnbull, Stuart},
	journal={Management Science},
	volume={57},
	number={7},
	pages={1267--1287},
	year={2011},
	publisher={INFORMS}
}

@article{tian2015variable,
	title={Variable selection and corporate bankruptcy forecasts},
	author={Tian, Shaonan and Yu, Yan and Guo, Hui},
	journal={Journal of Banking \& Finance},
	volume={52},
	pages={89--100},
	year={2015},
	publisher={Elsevier}
}

@article{loughran2011liability,
  title={When is a liability not a liability? Textual analysis, dictionaries, and 10-Ks},
  author={Loughran, Tim and McDonald, Bill},
  journal={The Journal of finance},
  volume={66},
  number={1},
  pages={35--65},
  year={2011},
  publisher={Wiley Online Library}
}

@article{gunning1952technique,
  title={Technique of clear writing},
  author={Gunning, Robert and others},
  year={1952},
  publisher={McGraw-Hill}
}

@article{araci2019finbert,
  title={Finbert: Financial sentiment analysis with pre-trained language models},
  author={Araci, Dogu},
  journal={arXiv preprint arXiv:1908.10063},
  year={2019}
}

@article{devlin2018bert,
  title={Bert: Pre-training of deep bidirectional transformers for language understanding},
  author={Devlin, Jacob and Chang, Ming-Wei and Lee, Kenton and Toutanova, Kristina},
  journal={arXiv preprint arXiv:1810.04805},
  year={2018}
}

@inproceedings{nguyen2021multimodal,
  title={Multimodal Machine Learning for Credit Modeling},
  author={Nguyen, Cuong V and Das, Sanjiv R and He, John and Yue, Shenghua and Hanumaiah, Vinay and Ragot, Xavier and Zhang, Li},
  booktitle={2021 IEEE 45th Annual Computers, Software, and Applications Conference (COMPSAC)},
  pages={1754--1759},
  year={2021},
  organization={IEEE}
}

@article{tetlock2008more,
  title={More than words: Quantifying language to measure firms' fundamentals},
  author={Tetlock, Paul C and Saar-Tsechansky, Maytal and Macskassy, Sofus},
  journal={The journal of finance},
  volume={63},
  number={3},
  pages={1437--1467},
  year={2008},
  publisher={Wiley Online Library}
}

@article{mayew2015md,
  title={MD\&A Disclosure and the Firm's Ability to Continue as a Going Concern},
  author={Mayew, William J and Sethuraman, Mani and Venkatachalam, Mohan},
  journal={The Accounting Review},
  volume={90},
  number={4},
  pages={1621--1651},
  year={2015},
  publisher={American Accounting Association}
}

@article{chava2022measuring,
  title={Measuring Firm-Level Inflation Exposure: A Deep Learning Approach},
  author={Chava, Sudheer and Du, Wendi and Shah, Agam and Zeng, Linghang},
  journal={Available at SSRN 4228332},
  year={2022}
}

@article{chava2021managers,
  title={Do Managers Walk the Talk on Environmental and Social Issues?},
  author={Chava, Sudheer and Du, Wendi and Malakar, Baridhi},
  journal={Available at SSRN 3900814},
  year={2021}
}

@article{chava2020buzzwords,
  title={Buzzwords?},
  author={Chava, Sudheer and Du, Wendi and Paradkar, Nikhil},
  journal={Available at SSRN 3862645},
  year={2020}
}

@article{tibshirani1996regression,
  title={Regression shrinkage and selection via the lasso},
  author={Tibshirani, Robert},
  journal={Journal of the Royal Statistical Society: Series B (Methodological)},
  volume={58},
  number={1},
  pages={267--288},
  year={1996},
  publisher={Wiley Online Library}
}

@article{hoerl1970ridge,
  title={Ridge regression: Biased estimation for nonorthogonal problems},
  author={Hoerl, Arthur E and Kennard, Robert W},
  journal={Technometrics},
  volume={12},
  number={1},
  pages={55--67},
  year={1970},
  publisher={Taylor \& Francis}
}

@article{schmidhuber2015deep,
  title={Deep learning in neural networks: An overview},
  author={Schmidhuber, J{\"u}rgen},
  journal={Neural networks},
  volume={61},
  pages={85--117},
  year={2015},
  publisher={Elsevier}
}

@article{agarwal2008comparing,
	title={Comparing the performance of market-based and accounting-based bankruptcy prediction models},
	author={Agarwal, Vineet and Taffler, Richard},
	journal={Journal of Banking \& Finance},
	volume={32},
	number={8},
	pages={1541--1551},
	year={2008},
	publisher={Elsevier}
}

@article{makwana4163283get,
  title={How to Get Investment Grade Rating in the Age of Explainable Ai?},
  author={Makwana, Ravi and Bhatt, Dhruvil and Delwadia, Kirtan and Shah, Agam and Chaudhury, Bhaskar},
  year={2022},
  journal={Available at SSRN 4163283}
}

@article{duffie2009frailty,
  title={Frailty correlated default},
  author={Duffie, Darrell and Eckner, Andreas and Horel, Guillaume and Saita, Leandro},
  journal={The Journal of Finance},
  volume={64},
  number={5},
  pages={2089--2123},
  year={2009},
  publisher={Wiley Online Library}
}
